\documentclass[a4paper,12pt]{amsart}
\usepackage[left=1.2in, right=1.2in, , bottom = 1.2in, top = 1.2in]{geometry}
\usepackage[utf8]{inputenc}
\usepackage[all]{xy}
\usepackage{amssymb}
\usepackage{amsmath}
\usepackage{csquotes}
\usepackage{subfig}
\usepackage{enumerate}
\usepackage{graphicx}
\usepackage{enumerate}
\usepackage{nameref}
\usepackage{letltxmacro}
\makeatletter
\AtBeginDocument{%
  \@ifdefinable{\myorg@nameref}{%
    \LetLtxMacro\myorg@nameref\nameref
    \DeclareRobustCommand*{\nameref}[1]{%
      \emph{\myorg@nameref{#1}}%
    }%
  }%
}
\makeatother
\usepackage{amsaddr}
\usepackage{color}
\usepackage{mwe}
\usepackage{subfig}
\usepackage{appendix}
\usepackage{cite} 

\usepackage{soul}
\usepackage{xcolor}
\newcommand\independent{\protect\mathpalette{\protect\independenT}{\perp}}
\def\independenT#1#2{\mathrel{\rlap{$#1#2$}\mkern2mu{#1#2}}}

\usepackage{tikz}
\usetikzlibrary{arrows.meta,shapes}
\usetikzlibrary{arrows,shapes.arrows,shapes.geometric,shapes.multipart,
decorations.pathmorphing,positioning,shapes.swigs,}
\usepackage{setspace}
\newtheorem{proposition}{Proposition}

\newtheorem*{assumption}{Assumption}

\MakeOuterQuote{"}\EnableQuotes
\DeclareUnicodeCharacter{00A0}{ }

\title[]{Identification of vaccine effects when exposure status is unknown}
 \author{Mats J. Stensrud$^{1}$, Louisa H. Smith$^2$} \address{ \small $^1$ Department of Mathematics, Ecole Polytechnique Fédérale de Lausanne, Switzerland \\
 $^2$ Roux Institute, Northeastern University, Portland, Maine, USA
 }

\begin{document}
\maketitle
\thispagestyle{empty}

\pagenumbering{arabic}
\begin{abstract}
Results from randomized controlled trials (RCTs) help determine vaccination strategies and related public health policies. However, defining and identifying estimands that can guide policies in infectious disease settings is difficult, even in an RCT. The effects of vaccination critically depend on characteristics of the population of interest, such as the prevalence of infection, the number of vaccinated, and social behaviors. To mitigate the dependence on such characteristics, estimands, and study designs, that require conditioning or intervening on exposure to the infectious agent have been advocated. But a fundamental problem for both RCTs and observational studies is that exposure status is often unavailable or difficult to measure, which has made it impossible to apply existing methodology to study vaccine effects that account for exposure status. In this work, we present new results on this type of vaccine effects. Under plausible conditions, we show that point identification of certain relative effects is possible even when the exposure status is unknown. Furthermore, we derive sharp bounds on the corresponding absolute effects. We apply these results to estimate the effects of the ChAdOx1 nCoV-19 vaccine on SARS-CoV-2 disease (COVID-19) conditional on post-vaccine exposure to the virus, using data from a large RCT.
\end{abstract}

 \section*{Introduction}
\label{sec: intro}



Vaccines are one of the most important inventions in modern medicine \cite{greenwood2014contribution}. Justification for real-life vaccination strategies relies heavily on results from large-scale vaccine randomized controlled trials (RCTs). However, the nature of communicable disease means that defining and evaluating vaccine effects requires consideration of population characteristics such as the prevalence of current and prior infection, mixing patterns, and concurrent public health measures.

Policy-relevant estimands for vaccine trials have been discussed extensively (see Halloran et al. \cite{halloran2010design} for an overview), in particular in the context of the SARS-CoV-2 disease (COVID-19) pandemic \cite{mehrotra2021clinical,gilbert2021assessment,lipsitch2020understanding,lipsitch2021interpreting,kilpatrick2020estimands,follmann2021vaccine,patel2021evaluation}. However, as of yet, methods to study vaccine effects conditional on, or under interventions on, exposure to the infectious agent are rarely used. Here and henceforth, we use exposure to mean exposure to the disease agent, such as a virus, which is distinct from the treatment, such as a vaccine. 
A key problem is that exposure status is often difficult, or even impossible, to measure in practice \cite{halloran2010design,o2014estimating}. For example, Halloran and Struchiner \cite{halloran1995causal} write that measuring susceptibility to infection "might not be easy in practice and might indeed require considerable assumptions regarding who is infectious and when, how infectious the persons are, and who is exposing whom." 
Challenge trials, in which participants are intentionally exposed, are one option for controlling exposure status but involve serious ethical issues \cite{jamrozik2020covid,corey2020strategic,cohen2016studies}.

This article specifically targets effects that account for exposure status, even when it is unmeasured. We provide results on the interpretation and identification of causal effects of vaccines from RCTs and observational studies. The results include identification results for the causal effect of a vaccine on clinical outcomes, conditional on an \textit{unmeasured} exposure to the infectious agent. Specifically, we show that, under a plausible no effect on exposure assumption, the relative effect -- though not the absolute effect -- of the vaccine can be point-identified in an RCT. Furthermore, under the same assumption, we derive sharp bounds for the absolute effect. We clarify how these effects are related to existing estimands and notions of biological effects, and we give identification results on per-exposure effects \cite{o2014estimating,struchiner2007randomization}, a type of controlled direct effect, even when the exposure is unmeasured, as is often the case in practice.

The article is organized as follows. Section \nameref{sec: observed data structure} presents the data structure and the notation. Section \nameref{sec: causal parameters} provides definitions and interpretation of causal estimands. Section \nameref{sec: identification} contains results on the identification of causal estimands, including point identification results for the relative causal effect conditional on exposure, and partial identification results for the absolute causal effect conditional on exposure. Section \nameref{sec: id external data} presents results for point identification of absolute causal effects conditional on exposure when external data on exposure risk are available, and suggests a sensitivity analysis when external data are unavailable. Section \nameref{sec: time to event} extends the results to time-to-event outcomes, in a setting in which individuals can be censored due to loss to follow-up. Section \nameref{sec: estimation} describes how our new parameters can be estimated using existing estimators, even when the outcome is unmeasured.  Section \nameref{sec: example} implements the new results in a study of the ChAdOx1 nCoV-19 (Oxford) vaccine against COVID-19.

 \section*{Data Structure}
\label{sec: observed data structure}
Suppose that we have data from a randomized experiment with $n$ individuals who are assigned a binary treatment $A \in \{0,1\}$ at baseline, where $A=1$ indicates receiving vaccine and $A=0$ indicates placebo or other control. As is common in vaccine trial settings \cite{tsiatis2021estimating,halloran1996estimability}, we consider inference in a much larger population from which the trial participants are drawn, so that interactions among patients in the trial are negligible; thus, we suppose the individuals are iid and omit the $i$ subscript. Let $L$ be a vector of baseline covariates. To simplify the presentation, we suppose $L$ is discrete, but the results generalize to continuous $L$. 

Let $E \in \{0,1\}$ be an indicator of whether an individual is exposed to the infectious agent at least once, e.g., being in close contact with an actively contagious individual, which may be \textit{unobserved} in the study. While we will focus on settings where the exposure $E$ occurs after treatment $A$ is assigned, i.e.\ after baseline, our results also allow the (unobserved) exposure $E$ to be temporally ordered, and thus occur, before treatment $A$. We first consider $Y \in \mathbb{R}_{\geq 0}$ to be the outcome of interest, e.g., disease severity or hospitalization, measured at a given time after randomization, where we define $Y = 0$ when an individual does not have the outcome. In Section \nameref{sec: time to event},  we extend the results to censored time-to-event outcomes. 

We use superscripts to denote counterfactuals \cite{robins1986new,richardson2013single}. For example, $Y^{a=1}$ and $Y^{a=0}$ are the outcomes of interest when the treatment is, possibly contrary to fact, fixed to active vaccine ($a=1$) or control ($a = 0)$.


 \section*{Causal Parameters}
\label{sec: causal parameters}

\subsection*{The average treatment effect (ATE)} 
\label{sec: ate}
To motivate the new contributions in this manuscript, we first review the conventional average treatment effect (ATE) of $A$ on the outcome $Y$,
\begin{align}
& \mathbb{E} ( Y^{a=1}) \text{ vs. }  \mathbb{E} ( Y^{a=0}),
\label{eq: ate}
\end{align}
which compares the average outcome in the trial population had everyone been treated ($a=1$) versus not treated ($a=0$). This contrast can be identified without additional assumptions when the trial is perfectly executed,  that is, under perfect randomization and no losses to follow-up. However, as with any trial, the magnitude of \eqref{eq: ate} depends on the specific setting in which the RCT was conducted; in a vaccine trial, crucial characteristics include the number of currently infected in the population, the number of previously infected, the mixing pattern, and additional public health measures that may be simultaneously implemented. To generalize the results from the RCT to a policy-relevant setting, we must account for these characteristics, which is far from straightforward. 

\subsubsection*{Conditional Counterfactual Contrasts Are Not Necessarily Causal Effects}
To mitigate some of the concerns that are raised about the ATE in vaccine trials, we could attempt to adjust for exposure to the infectious agent \cite{halloran2010design,halloran1995causal}. However, defining causal effects conditional on exposure status is not straightforward because exposure status is a post-treatment variable. In particular, a naive contrast of counterfactual outcomes conditional on exposure status,
\begin{align}
& \mathbb{E} (Y^{a=1} \mid E^{a=1}=1 ) \text{ vs. }  \mathbb{E} (Y^{a=0}  \mid E^{a=0}=1) \label{eq: naive contrast}, 
\end{align}
is not a causal effect when the treatment affects the post-treatment event; it compares counterfactual outcomes in different subpopulations of individuals. This is illustrated by the path $A \rightarrow E$ in the causal directed acyclic graph DAG in Figure \ref{fig: swigs}a, which leads to an indirect effect of vaccination on the outcome $Y$ through the path $A \rightarrow E \rightarrow Y$. This indirect effect is plausible if participants know their treatment status; for example, one may expect that vaccinated individuals show a reduction in protective behaviours, which increases the risk of being exposed.

\subsection*{The Principal Stratum Effect (PSE) and the Causal Effect Conditional on Exposure (CECE)}
\label{sec: principal stratum eff}
A principal stratum effect (PSE) \cite{robins1986new,frangakis2002principal} compares counterfactual outcomes among individuals with the same counterfactual exposure status. We can define a particular PSE among those individuals who would be exposed to the infectious agent, at least once, regardless of treatment assignment,
\begin{align}
& \mathbb{E} (Y^{a=1} \mid E^{a=0}=E^{a=1}=1) \text{ vs. }  \mathbb{E} (Y^{a=0}  \mid E^{a=0}=E^{a=1}=1). \label{eq: ps effect}
\end{align}
Unlike \eqref{eq: naive contrast}, the PSE \eqref{eq: ps effect} is a contrast of counterfactual outcomes in the same (sub)population of individuals, and it is therefore a causal effect. However, the conditioning set in \eqref{eq: ps effect} is defined by exposures in the same individual under two different treatments and, without further assumptions, it is impossible to observe the individuals in this subpopulation \cite{robins1986new}, even when $E$ is measured. Thus, the PSE is defined in an unknown subpopulation that is unobservable even in principle, and the practical relevance of the PSE has been seriously questioned \cite{ robins2007principal, joffe2011principal, dawid2012imagine,vanderweele2011principal}. 

As an alternative to the PSE, consider a contrast of counterfactual outcomes conditional on exposure status in the observed data,
\begin{align}
    \mathbb{E} (Y^{a=1} \mid E = 1 ) & \text{ vs. } \mathbb{E} (Y^{a=0} \mid E =  1 ). \label{eq: cdce}
\end{align}

Like \eqref{eq: ps effect}, the contrast in \eqref{eq: cdce} is a causal effect as it compares the same subpopulation of individuals under different treatment. Unlike \eqref{eq: ps effect}, the conditioning set in \eqref{eq: cdce} is observable when $E$ is measured. Without additional assumptions, however, the interpretation of \eqref{eq: cdce} is not straightforward, because an individual's exposure status in the observed world ($E$) is not guaranteed to be equal to the exposure status under an intervention that fixes the treatment to be $a$ ($E^a$). Thus, in general we cannot interpret \eqref{eq: cdce} as a direct effect of treatment $A$ on the outcome $Y$ outside of the treatment effects on exposure status. 

But there is at least one setting in which differences in exposure status would not be expected between treatment groups: a \textit{blinded} RCT, which is the context of many vaccine efficacy studies. The following mechanistic assumption formalizes the notion that receiving the vaccine does not exert effects on exposure status $E$.

\begin{assumption}[No effect on exposure]
\begin{align}
E^{a=0} =  E^{a=1}  \label{ass: no indirect vaccine effect}.
\end{align} 
\end{assumption}

Assumption \eqref{ass: no indirect vaccine effect} guarantees that exposure to the infectious agent is the same, regardless of the treatment that was assigned, and, assuming that the intervention on $A$ is well-defined, allows us to write $E = E^{a=0} =  E^{a=1} $. 

This assumption can also hold outside of a blinded RCT. In particular, exposures that are outside of the individual's control can satisfy Assumption \eqref{ass: no indirect vaccine effect}. Such exposures could be consequences of natural or human disasters, such as a flooding after an intense rainfall or radiation from an atomic bombing.

The DAG in Figure \ref{fig: swigs}b describes the causal structure of a blinded RCT, in which this assumption would be expected to be met, as there is no path $A \rightarrow E$ and therefore no indirect effect of vaccination on the outcome through the path $A \rightarrow E \rightarrow Y$.  

Under assumption \eqref{ass: no indirect vaccine effect}, the contrasts \eqref{eq: naive contrast}-\eqref{eq: cdce} are equal, that is,
\begin{align}
  \mathbb{E} (Y^{a=1} \mid E = 1 ) & \text{ vs. } \mathbb{E} (Y^{a=0} \mid E =  1 ) \nonumber \\
 = \mathbb{E} (Y^{a=1} \mid E^{a=1} = 1 ) & \text{ vs. } \mathbb{E} (Y^{a=0} \mid E^{a=0} =  1 ) \nonumber \\
=  \mathbb{E} (Y^{a=1} \mid E^{a=0} = E^{a=1} = 1 ) & \text{ vs. } \mathbb{E} (Y^{a=0} \mid E^{a=0} = E^{a=1} = 1 ). \nonumber 
\end{align}

Halloran and Struchiner \cite{halloran1995causal} also advocated contrasts of (counterfactual) outcomes in exposed individuals, under the assumption that "people did not change their behavior after randomization"\cite{halloran1995causal}[Page 147]. Condition \eqref{ass: no indirect vaccine effect} formalizes when such contrasts are unambiguous causal effects, i.e.\ contrast of outcomes in the same (sub)population of individuals.

Because we focus on blinded RCTs in this work, we will use assumption \eqref{ass: no indirect vaccine effect} extensively, and under \eqref{ass: no indirect vaccine effect} we will denote the contrasts \eqref{eq: ps effect}-\eqref{eq: cdce} collectively as the causal effect conditional on exposure (CECE), which is also equal to \eqref{eq: naive contrast}.

The CECE mitigates some of the concerns that are raised about the generalizability of the ATE \eqref{eq: ate}, because the CECE is confined to those individuals who are exposed to the infectious agent in the observed data, regardless of treatment assignment. Thus, assumption \eqref{ass: no indirect vaccine effect} ensures that the CECE has a mechanistic interpretation as an average causal effect given exposure to the infectious agent. The CECE is also of immediate interest for individuals who, based on their own subject-matter knowledge, believe, or possibly know, that they will be, or already are, exposed. 

However, the CECE is defined among those who would be exposed in a given study, and the subset who is exposed is context-dependent. To understand the CECE, it is helpful to draw an analogy to ring vaccination trials, in which individuals are recruited only if they have been exposed to an index case, and are subsequently randomly assigned to $A$. Suppose we indicate exposure to an index case as $E$. Then, the estimand \eqref{eq: cdce} corresponds to the usual estimand in ring vaccination trials, which is an effect conditional on being exposed. Like the CECE, the target population of a ring vaccination trial is context-dependent, as being a contact of an index case is required for inclusion, and characteristics of those individuals depend on the setting. In a ring vaccination trial, however, exposure is pre-treatment and exposure status is known, features not shared with our setting.

\subsection*{The Controlled Direct Effect (CDE)}
\label{sec: cde}
A special case of a controlled direct effect (CDE) \cite{robins1992identifiability}, also called a per-exposure effect or a challenge effect \cite{o2014estimating,halloran1995causal}, is defined with respect to an intervention on the treatment $A$ and the exposure $E$,
\begin{align}
& \mathbb{E} (Y^{a=1,e=1}) \text{ vs. }  \mathbb{E} (Y^{a=0,e=1}).
\label{eq: cde}
\end{align}
This CDE corresponds to the effect that is identified by a challenge trial \cite{hudgens2009assessing}; that is, a study where the participants are subject to an intervention where they are guaranteed to be physically exposed to the infectious agent. Outside of RCTs, household studies are sometimes used to infer such effects, based on contrasts of household secondary attack rates \cite{halloran1995causal}. 

Unlike the ATE \eqref{eq: ate}, the CDE is defined in a controlled setting, in which \textit{all} individuals are exposed to the infectious agent. Thus, this effect is insensitive to the risk of exposure in the observed population.

Finally, all the estimands considered in this section can be defined conditional on any baseline covariate $L$. The distinction between estimands conditional on $L$ and marginal estimands will be of interest when we study identification in Section \nameref{sec: identification}.

\subsection*{The notion of a "biological" effect}
Both the CECE and CDE quantify treatment effects in individual who are guaranteed to be exposed to the disease agent. In that sense, both effects seem to be captured by the notion of "biological" effects. However, the fact that the CECE and CDE are distinct estimands illustrates that the term "biological" effect, without further clarification, is ambiguous.

\section*{Identification} 
\label{sec: identification}
To motivate the identification results in this work, we first review three standard identifiability conditions for the ATE.
\begin{assumption}[Treatment exchangeability]
\begin{align}
& Y^a, E^a \independent A \label{ass: cond trt ech} \ \forall a \in \{0,1\}.
\end{align} 
 \end{assumption}
Treatment exchangeability e.g. holds in the Single World Intervention Graph (SWIG) \cite{richardson2013single} in Figure \ref{fig: swigs}c, even if $L$ is unmeasured. 
\begin{assumption}[Positivity]
\begin{align}
& P(A =a ) > 0 \quad  \forall a \in \{0,1\}.  \label{ass: pos}
\end{align} 
 \end{assumption}
 
 \begin{assumption}[Consistency]
\begin{align}
& \text{ If }    A=a,  \text{ then } {E} = {E}^{a},  {Y} = {Y}^{a} \ \forall a \in \{0,1\}.  \label{ass: consistency}
\end{align} 
 \end{assumption}

Conditions \eqref{ass: cond trt ech}-\eqref{ass: consistency} hold by design in an RCT where treatment is unconditionally randomly assigned. These three conditions allow us to identify the ATE \eqref{eq: ate} as 
$ \mathbb{E}(Y \mid A = 1) \text{ vs. } \mathbb{E}(Y \mid A = 0)$, regardless of whether exposure status $E$ is measured. 

However, our focus is on estimand \eqref{eq: cdce} (and \eqref{eq: cde} in e\ref{sec: id cde}), which is defined with respect to counterfactual statuses of the exposure $E$, so which require additional assumptions.
\subsection*{Identification of the CECE}
\label{sec: id cdce}


Under the no effect on exposure assumption \eqref{ass: no indirect vaccine effect} and conditions \eqref{ass: cond trt ech}-\eqref{ass: consistency}, it is straightforward to express the CECE as a function of factual variables, 
\begin{align}
    \mathbb{E} (Y^{a=1} \mid E = 1 ) & \text{ vs. } \mathbb{E} (Y^{a=0} \mid E =  1 ) \nonumber \\
    = \mathbb{E} (Y \mid E = 1, A = 1 ) & \text{ vs. } 
    \mathbb{E} (Y \mid E = 1, A = 0 ), \nonumber 
\end{align}
but the CECE, as defined as an arbitrary contrast ("vs."), is not point identified in our data because  $\mathbb{E} (Y \mid E = 1, A = a) $ is not estimable when $E$ is unmeasured. For example, the absolute CECE,
$$
    \mathbb{E} (Y^{a=1} \mid E=1) -  \mathbb{E} (Y^{a=0}  \mid E=1)  = \mathbb{E} (Y \mid E = 1, A = 1) - \mathbb{E} (Y \mid E = 1, A = 0),
$$
is not possible to estimate from the observed data. 

To identify the CECE, we therefore introduce an additional assumption, which relates the unmeasured $E$ to $Y$.
\begin{assumption}[Exposure necessity]
 \begin{align}
     E^a = 0 \implies Y^a = 0, \ \forall a \in \{0,1\}. \label{ass: exposure necessity}
 \end{align}
\end{assumption}
The exposure necessity assumption states that only individuals who were exposed to the infectious agent can experience the outcome. Thus, the exposure is a necessary condition for experiencing the outcome.  For example, contact with some amount of live virus is necessary to develop severe disease. Many exposures and outcomes of interest meet this criterion, though sometimes researchers may be interested in other exposures that do not necessarily satisfy this criterion, e.g., sharing a home or classroom with an infected individual. However, such an exposure definition might be revised to being in the same room with an infected individual for at least 1 minute, though even that might not be strictly necessary. In practice, it is important that the investigator has articulated a well-defined exposure, but it is possible that different investigators use different definitions. 

Our first proposition shows that the \textit{relative} CECE is identified under the conditions we have introduced so far, which are expected to hold in a blinded RCT. 

\begin{proposition}[Relative CECE]
\label{thm: relative cdc}
Under the no effect on exposure assumption \eqref{ass: no indirect vaccine effect}, standard identifiability conditions \eqref{ass: cond trt ech}-\eqref{ass: consistency} and exposure necessity \eqref{ass: exposure necessity}, the relative CECE is equal to 
\begin{align}
    &  \frac{\mathbb{E} (Y^{a=1} \mid E=1)}{ \mathbb{E} (Y^{a=0}  \mid E=1)} 
    = \frac{ \mathbb{E} (Y \mid A = 1)}{ \mathbb{E} (Y \mid A = 0)}, \nonumber 
\end{align}
given that $\mathbb{E}(Y  \mid A = 0) > 0$. 
\end{proposition}

The proof is found in e\ref{app sec: proofs}. From our considerations in Section \nameref{sec: principal stratum eff} and our derivations in Section \nameref{sec: id cdce},  it follows that Proposition \ref{thm: relative cdc} also gives an identification result for the relative principal stratum effect, that is,  $     \frac{\mathbb{E} (Y^{a=1}  \mid E^{a=0} = E^{a=1} = 1 ) }{\mathbb{E} (Y^{a=0}  \mid E^{a=0} = E^{a=1} = 1 )} $. Interestingly, Proposition \ref{thm: relative cdc} shows that the relative CECE is equal to the conventional ATE on the relative risk scale, which is routinely reported in RCTs. Thus, we specify the  assumptions that allow for interpretation of this estimand as a measure of vaccine efficacy conditional on exposure to infection \cite{halloran2010design}. 

The fact that the relative CECE is identified by the same formula as the relative ATE is related to the known result in epidemiology that diagnostic tests that have perfect specificity will give unbiased estimates of risk ratios, even if these tests do mis-classify disease cases. We discuss this in e\ref{sec app:specificity}.  

Whereas the absolute CECE is not point identified, our next proposition gives partial identification of the absolute CECE for a binary outcome $Y \in [0,1]$ in terms of sharp bounds. To simplify the presentation of the subsequent results we suppose, without loss of generality, that $\mathbb{E} (Y \mid A = 0)  \geq \mathbb{E} (Y \mid A = 1) $.

\begin{proposition}[Absolute CECE]
\label{thm: absolute cdc}
Under the no effect on exposure assumption \eqref{ass: no indirect vaccine effect} and conditions \eqref{ass: cond trt ech}-\eqref{ass: exposure necessity}, the absolute CECE on an outcome $Y \in [0,1]$ is partially identified by the sharp bounds 
\begin{align}
    & {\mathbb{E} (Y \mid  A = 0)   }  - {\mathbb{E} (Y \mid A = 1) } \leq  {\mathbb{E} (Y^{a=0}  \mid E  = 1 ) } - {\mathbb{E} (Y^{a=1}  \mid E= 1 )}  \leq 1 - \frac{ \mathbb{E} (Y \mid A = 1)}{ \mathbb{E} (Y \mid A = 0)}, \nonumber 
\end{align}
when $\mathbb{E} (Y \mid A = 0)  \geq \mathbb{E} (Y \mid A = 1) $.
\end{proposition}
The proof is given in e\ref{app sec: proofs}.

\subsection*{Remark on Proposition \ref{thm: absolute cdc}}
The lower bound on the absolute CECE is equal to the absolute ATE. Thus, Proposition \ref{thm: absolute cdc} gives us another interpretation of a standard risk difference -- as a lower bound on the absolute CECE. Furthermore, this lower bound is equal to the absolute CECE if \textit{everybody} is exposed.

The upper bound is 1 minus the relative ATE, which is a quantity that is often reported as the vaccine efficacy in randomized controlled trials \cite{halloran2010design}, e.g.\ during the COVID-19 pandemic \cite{kahn2021leveraging}. The absolute CECE is equal to this bound if an unvaccinated individual ($A=0$) will experience the outcome ($Y=1$) if and only if she is exposed ($E=1$).  

It follows from Proposition \ref{thm: absolute cdc} that the larger $\mathbb{E} (Y \mid  A = 0) $, the more informative are the bounds. In particular, the lower bound is equal to the upper bound when $\mathbb{E} (Y \mid  A = 0) = 1$.

Zhao et al \cite{zhao2020note} studied another interesting setting where relative -- but not absolute -- risks could be point identified. Their causal question, which concerned racial discrimination in policing, was studied in a setting where the treatment, equivalent to our $A$, was unmeasured, but the mediator, equivalent to our $E$, was measured. Their estimand of interest was the conventional ATE.





\section*{External Data and Sensitivity Analysis}
\label{sec: id external data}
Consider a binary outcome $Y \in \{0,1\}$, e.g., an indicator of symptomatic disease. Suppose that the investigator has external knowledge about the risk of experiencing the outcome given exposure among the unvaccinated, that is, $P(Y = 1 \mid E = 1, A = 0) $. Alternatively, suppose that the investigator has external knowledge about the risk of being exposed among the unvaccinated, that is, $P(E=1 \mid A = 0)$. Knowledge of either of these probabilities could have been collected among trial eligible individuals who did not participate in the randomized experiment, or among a subset of the trial participants. 

Our next proposition shows that knowledge of either $P(Y = 1 \mid E = 1, A = 0) $ or $P(E=1 \mid A = 0)$  allows point identification of the absolute CECE, when we also assume the same identification conditions as in Proposition \ref{thm: absolute cdc}.

\begin{proposition}[Point identification of the absolute CECE]
\label{thm: sensitivity motivation}
Under the no effect on exposure assumption \eqref{ass: no indirect vaccine effect} and conditions \eqref{ass: cond trt ech}-\eqref{ass: exposure necessity}, 
\begin{align}
& {\mathbb{E} (Y^{a=0}  \mid E  = 1 ) } - {\mathbb{E} (Y^{a=1}  \mid E = 1 )} \nonumber \\
= &  \mathbb{E} (Y  \mid E = 1, A = 0)  \left( 1 - \frac{\mathbb{E} (Y \mid A = 1)  }{ \mathbb{E} (Y \mid A = 0)  } \right) \label{eq: thm external 1} \\
= &  \frac{\mathbb{E} (Y \mid  A = 0)   }{ P(E = 1 \mid A = 0)} -  \frac{\mathbb{E} (Y \mid A = 1)   }{ P(E = 1 \mid A = 1)  } \label{eq: thm external 2} .
\end{align}
\end{proposition}
The proof of Proposition \ref{thm: sensitivity motivation} is given in e\ref{sec app: sensitivity analysis}. Besides giving point identification results in settings with knowledge from external data, Proposition \ref{thm: sensitivity motivation} motivates sensitivity analyses for the magnitude of the absolute CECE using sensitivity parameters that are justified by subject-matter reasoning; that is, the investigator can evaluate \eqref{eq: thm external 1} and \eqref{eq: thm external 2} under different values of the marginal sensitivity parameters $P(Y = 1 \mid E = 1, A = 0) $ and $P(E=1 \mid A = 0)$, respectively.

\section*{CECE in Time-to-Event Settings}
\label{sec: time to event}
In both RCTs and observational studies, it is common to evaluate vaccine effects on time-to-event outcomes. Our results generalize to settings where the exposure status and the outcome of interest are both time-to-event variables, which possibly are censored due to losses to follow-up. 

Suppose that $Y_k$ and $E_k$ are time-to-event variables indicating whether an individual has experienced the event by time $k$, i.e., $Y_k = 1$, and has been exposed by time $k$, respectively. That is, $E_k = 1$ means exposure has occurred at least once. Let $C_{k}$ indicate loss to follow-up (censoring) by time $k > 0$. To align with the established causal inference literature \cite{hernan2018causal,robins1986new,richardson2013single}, suppose that we are interested in outcomes in discrete time intervals $k=0,\dots K$, and define the temporal (and topological) order $(C_{k},E_{k},Y_{k})$ in each interval $k > 0$. This setting will converge to a continuous time setting when we let the time intervals become small. We continue to use superscripts to denote counterfactuals, and we formally consider a counterfactual estimand under interventions on the baseline treatment $A$ and the censoring variable $C_k$ \cite{robins2000correcting,young2018choice}. For example, $Y_k^{a,c=0}$ is the counterfactual outcome of interest by time $k$ when treatment is assigned to $a$ and there is no loss to follow-up. The Single World Intervention Graph (SWIG) in Figure \ref{fig:time to event} describes a causal structure that is coherent with our time-to-event setting. In e\ref{sec app: time-to-event}, we give more details on the time-to-event notation, and we state generalizations of the exchangeability, positivity, consistency, exposure necessity and the no effect on exposure conditions to settings with time-to-event outcomes, see conditions \eqref{ass: time ech}--\eqref{ass no exposure eff time-var}. Under these conditions, we can identify the relative CECE as a ratio of cumulative incidences, as described in the next proposition.

\begin{proposition}[Relative and absolute CECE for time to event outcomes]
\label{thm: rel CECE time var}

Under exchangeability, positivity, consistency, exposure necessity and the no effect on exposure assumption for time-to-event outcomes, formally stated as conditions \eqref{ass: time ech}-\eqref{ass no exposure eff time-var} in e\ref{sec app: time-to-event}, the relative CECE at time $k$, $0\leq k \leq K$, is identified by the ratio of cumulative incidences, 
\begin{align}
    & \frac{\mathbb{E} (Y_{k}^{a=1,c=0}  \mid E_{k}^{a=1,c=0}  = 1 ) }{\mathbb{E} (Y_{k}^{a=0,c=0}  \mid E_{k}^{a=0,c=0}= 1 )}  = \frac{ \mu_k(1) }{ \mu_k(0)}, \nonumber 
\end{align}
where 
$$
\mu_k(a) = \sum_{s=1}^{k}h_{s}(a)\prod_{j=0}^{s-1}\left[1-h_{j}(a)\right]
$$
and
$$
h_{k}(a)=\frac{\mathbb{E}\lbrack Y_{k}(1-Y_{k-1}) (1-C_k) \mid A=a]}{%
\mathbb{E}\lbrack (1-Y_{k-1})(1-C_k) \mid A=a]}.  $$
Under the same conditions, the absolute CECE is partially identified by the sharp bounds \begin{align}
    & \mu_k(0) - \mu_k(1)  \leq  {\mathbb{E} (Y_{k}^{a=0,c=0}   \mid E_{k}^{a=0,c=0}   = 1 ) } - \mathbb{E} (Y_{k}^{a=1,c=0}   \mid E_{k}^{a=1,c=0}=1 ) \leq 1 - \frac {\mu_k(1) }{  \mu_k(0) }, \nonumber 
\end{align}
when $\mu_k(0)  \geq \mu_k(1) $.
\end{proposition}

See e\ref{sec app: time-to-event} for details and a proof. Thus, like the point exposure and point outcome setting, we do not need to measure common causes of $E_j$ and $Y_k$, $j,k \in \{0,\dots,K\}$, in order to point identify the relative CECE and bound the absolute CECE in time-to-event settings.

We have restricted all our discussion to results on \textit{risks}, not \textit{rates} such as hazards. Despite the fact that hazards are sometimes reported as "efficacy parameters" in infectious disease settings, there are well-known limitations of considering causal estimands on the hazard scale, see e.g.\ \cite{robins1986new,hernan2010hazards,stensrud2019limitations,stensrud2020test}, because of the conditioning on a post-treatment event -- here outcomes at earlier times -- that is affected by treatment. 

\subsection*{Excess and Etiologic Fractions}
\label{sec: attributable frac}

Following Greenland and Robins \cite{greenland1988conceptual}, the excess (prevented) fraction quantifies the excess of outcomes under treatment vs.\ control. When the assumptions of Proposition \ref{thm: rel CECE time var} hold, the excess fraction among the exposed is
\begin{align}
   & \frac{\mathbb{E} (Y_{k}^{a=0,c=0}  \mid E_{k}^{a=0,c=0}  = 1 ) - \mathbb{E} (Y_{k}^{a=1,c=0}  \mid E_{k}^{a=1,c=0}  = 1 ) }{\mathbb{E} (Y_{k}^{a=0,c=0}  \mid E_{k}^{a=0,c=0}= 1 )}  = 1 - \frac{ \mu_k(1) }{ \mu_k(0)}, 
      \label{eq: excess fraction}
\end{align}
which quantifies the increase in caseload under no treatment \cite{greenland1988conceptual,robins1989estimability}. In particular, the excess fraction conditional on exposure is equal to the unconditional excess fraction. Furthermore, \eqref{eq: excess fraction} is often what is reported as \textit{the vaccine efficacy} in clinical studies \cite{halloran2010design}.

The excess fraction should not be confused with the etiologic fraction, which is the fraction \textit{caused} by treatment. For example, suppose we consider outcomes at time $k$, and there are no losses to follow-up. Consider an individual for whom a vaccine prolonged the time to the outcome of severe infection from time $j$ to time $l$, but (s)he would nevertheless have a severe infection by time $k$ when taking the vaccine, where $j < l < k$. Then, treatment $A$ was a contributory cause of the outcome in this individual, and would thus count as an etiologic event in the etiologic fraction. On the other hand, the individual would not increase the excess caseload by time $k$, because (s)he experienced the outcome by time $k$ regardless of treatment. The etiologic fraction requires much stronger conditions for identification, even in RCTs and even without conditioning on exposure \cite{greenland1988conceptual,robins1989estimability}. 

\section*{Estimation and Implementation}
\label{sec: estimation}
Because all our identifying formulas from  Section \nameref{sec: identification}  are expressed in terms of simple conditional means, we can use simple estimators with known properties. Let $\hat{\mu}({a})$ and $\hat{\mu}({a,l})$ be estimators of $\mathbb{E}(Y \mid A =a ) $ and $\mathbb{E}(Y \mid A =a, L = l ) $, respectively, e.g.\ empirical means. We can estimate the relative CECE by
$$\widehat{\text{rCECE}} = \frac{\hat{\mu}({1})}{\hat{\mu}({0})}, $$ 
and similarly the upper bound on the absolute CECE by $\widehat{\text{aCECE}}_U  = 1 - \widehat{\text{rCECE}} $, where we can compute confidence intervals using standard estimators for risk ratios. Estimators of confidence intervals for risk ratios could be derived from Fieller's theorem \cite{fieller1954some} or the Delta method \cite{herson1975fieller}. The estimator of the lower bound on the absolute CECE is $\widehat{\text{aCECE}}_L  = \hat{\mu}({0}) - \hat{\mu}({1)}  $, which is simply a difference in means estimator. The estimator for the relative conditional CDE is defined analogously to $\widehat{\text{rCECE}}$, where we also include $L$ in the conditioning set, that is, $\widehat{\text{rCDE}} = {\hat{\mu}({1, l})}/{\hat{\mu}({0, l})}$.

For the identifying formulas in Section \nameref{sec: time to event}, which are cumulative incidences, let $\hat{\mu}_k({a})$ and $\hat{\mu}_k({a,l})$ be estimators of ${\mu}_k({a})$ and ${\mu}_k({a,l})$, respectively. Then, we can follow standard procedures for calculating ratios of cumulative incidence functions with confidence intervals, see e.g.\ \cite{zhang2008summarizing}[Sections 2.3 and 2.4].


\section*{Example: Effects of COVID-19 Vaccination}
\label{sec: example}


To study the effect of the ChAdOx1 nCoV-19 vaccine against COVID-19, Voysey et al \cite{voysey2021safety} enrolled 23848 participants in a blinded RCT done across the UK, Brazil, and South Africa. The participants were randomly assigned 1:1 to the ChAdOx1 nCoV-19 vaccine or control, which contained a meningococcal vaccine. The interim analysis included 11636 participants \cite{voysey2021safety}. The cumulative incidence of COVID-19 80 days since second dose was $0.9 \% \  (95\% \ \text{CI:} \ 0.5\%-1.3\%)$ and $3.1 \%  \ (95\% \  \text{CI}: 2.4\%-3.8\%)$ in the vaccine and placebo arms, respectively. Thus, an estimate of the relative $\text{CECE} \equiv \text{CECE}_{k=80}$, defined on the cumulative incidence scale, is
$$
\widehat{\text{rCECE}} =  \frac{ \hat{\mu}(1)}{ \hat{\mu}(0)} = 0.30 \ (95\% \ \text{CI:} \ 0.15-0.44),
$$
which corresponds to the reported vaccine efficacy point estimate of $1-0.30 = 0.70$ \cite{voysey2021safety}[Table 2]. Here and henceforth we omit the $k=80$ subscript to simplify the notation. 

We can use the results from Section \nameref{sec: id cdce} to derive bounds for the absolute CECE, specifically the sharp lower bound 
$$
\widehat{\text{aCECE}}_L = 0.031 - 0.009 = 0.022 \ (95\% \ \text{CI:} \ 0.011-0.033),
$$
and the sharp upper bound 
$$
\widehat{\text{aCECE}}_U = 1 - 0.30 = 0.70 \ (95\% \ \text{CI:} \ 0.57-0.85).
$$
Although we obtained informative point estimates of the relative CECE, the bounds on the absolute CECE are wide. The fact that the bounds are wide is not surprising, because they crucially depend on the risk of exposure to the virus. As discussed in Section \nameref{sec: id cdce}, the lower bound is reached under a setting where everybody is exposed to the virus, and the upper bound when the probability of the outcome among the exposed, unvaccinated individuals is 1. Depending on the definition of exposure, such settings may or may not be plausible. However, we can use a sensitivity analysis, as suggested in Section \nameref{sec: id external data}, to reason about the magnitude of the absolute CECE.

\subsection*{Sensitivity analysis in the  ChAdOx1 nCoV-19 vaccine study}
Determining sensitivity parameters to generate point estimates of the absolute CECE requires us to think concretely about the definition of exposure, or to consider a range of definitions of exposure. The ChAdOx1 nCoV-19 vaccine trial began enrollment in June 2020 and recruited a sample of 60-90\% health-care workers, depending on the site. Suppose we define $E$ as coming into contact with an equivalent amount of SARS-CoV-2 virus particles that may be encountered while caring for a COVID patient wearing personal protective equipment (PPE). However, because it is important that our definition of $E$ satisfies exposure necessity, we can more precisely define $E$ as a specific amount of virus particles such that the exposure necessity condition holds. For example, this could be a particular amount of virus particles when wearing PPE, and a higher amount when not wearing PPE.  To parameterize $P(E = 1 \mid A = 0)$, we might propose that 60\% of the trial participants were exposed to such an amount of virus particles at some point during the 80-days period. Given the observed data, this would imply that $P(Y = 1 \mid  E = 1, A = 0) = 0.052$; that is, $E$, here denoting a given amount of virus particles, was sufficient to cause symptomatic COVID-19 in just over 5\% of unvaccinated participants during 80 days of follow-up. In this setting, we would estimate $\widehat{\text{aCECE}} = 0.037$ (Figure \ref{fig:sensitivity parameters}). Suppose now that we rather set the sensitivity parameter $P(E = 1 \mid A = 0)$ to 0.9 instead of 0.6. Then, $\widehat{\text{aCECE}} = 0.024$, which is consistent with $P(Y = 1 \mid  E = 1, A = 0) = 0.034$. 

So far we have reasoned about the sensitivity parameter $P(E = 1 \mid A = 0)$. However, we could also reason about $P(Y = 1 \mid  E = 1, A = 0)$, perhaps using external data.  For example, consider the choir practice in Washington state in March 2020, after which 52 out of 61 participants developed COVID-19, having been exposed to a high concentration of virus in an unmasked setting \cite{hamner2020high}. Using this estimate of $P(Y = 1 \mid  E = 1, A = 0) = 0.85$, we find that $\widehat{\text{aCECE}} = 0.60$, an estimate consistent with $P(E = 1 \mid A = 0) = 0.036$. If exposure to such a high dose of SARS-CoV-2 is necessary for infection, the absolute CECE is much closer to its upper bound, and the risk of such an exposure in the trial setting is necessarily lower.

More broadly, the bounds illustrate an important point: the relative CECE is constant for \textit{any} exposure when Assumptions \eqref{ass: exposure necessity} and \eqref{ass: no indirect vaccine effect} hold, but only weak conclusions can be made about the magnitude of the absolute CECE unless we both have a clear idea about the definition of the exposure and have information about $P(Y = 1 \mid  E = 1, A = 0) $ or $P(E = 1 \mid A = 0)$.

Estimating the CDE requires data on covariates $L$, such as comorbidities, smoking, work occupation and age,  to justify condition \eqref{ass: cond exp ech}, see e\ref{sec: id cde}. Because this information is unavailable, we have not attempted to estimate the CDE.

 \section*{Discussion}
\label{sec: discussion}



In this work we have distinguished various estimands for vaccine effects conditional on exposure to infection and clarified their identification assumptions. We have required that the exposure, e.g., close contact with an infectious individual, is necessary for the outcome of interest to occur, e.g., symptomatic disease, as stated in our exposure necessity condition \eqref{ass: exposure necessity}. An alternative approach would involve adapting the definition of exposure to something that is possible to measure. For example, one might define exposure as close contact with infected people who present overt disease. However, such definitions have explicitly been discouraged, precisely because they would lead to an \textit{underestimate} of the exposure in settings where some infections are inapparent \cite{halloran2010design}. In the case of the CECE, we have considered the exposure to be any event such that the exposure necessity condition holds. 



When a necessary exposure is unmeasured, we have shown that relative effects can be point identified under plausible conditions, but absolute effects can only be bounded under the same conditions. Often the most commonly reported and publicized results are relative effects, as in major studies on different COVID-19 vaccines \cite{baden2021efficacy,thomas2021safety,voysey2021safety}. Thus, the results presented in this work give valuable interpretations to the numbers that are computed.  

However, often both relative and absolute effects are of interest. Absolute effects are usually studied in optimal regime settings \cite{murphy2003optimal,robins2004optimal,tsiatis2019dynamic}, which reflects the common opinion that heterogeneous effects on the additive scale are most appropriate for evaluating public health interventions \cite{knol2012recommendations}. Importantly, bounds on the additive effect can be used in formal decision theoretic approaches, even if these bounds are wide or cover null effects \cite{Cui2021Individualized,manski2018reasonable}. Furthermore, if the investigator is willing to invoke assumptions about the probability of exposure, the bounds will be narrower, as we describe in Section \nameref{sec: id external data}. 

In future work, we will formally consider generalizability of the different vaccine effects on different scales, including the CECE, which could be applicable to settings with interference outside of the randomized experiment. 




\bibliography{references}
\bibliographystyle{unsrt}

\clearpage

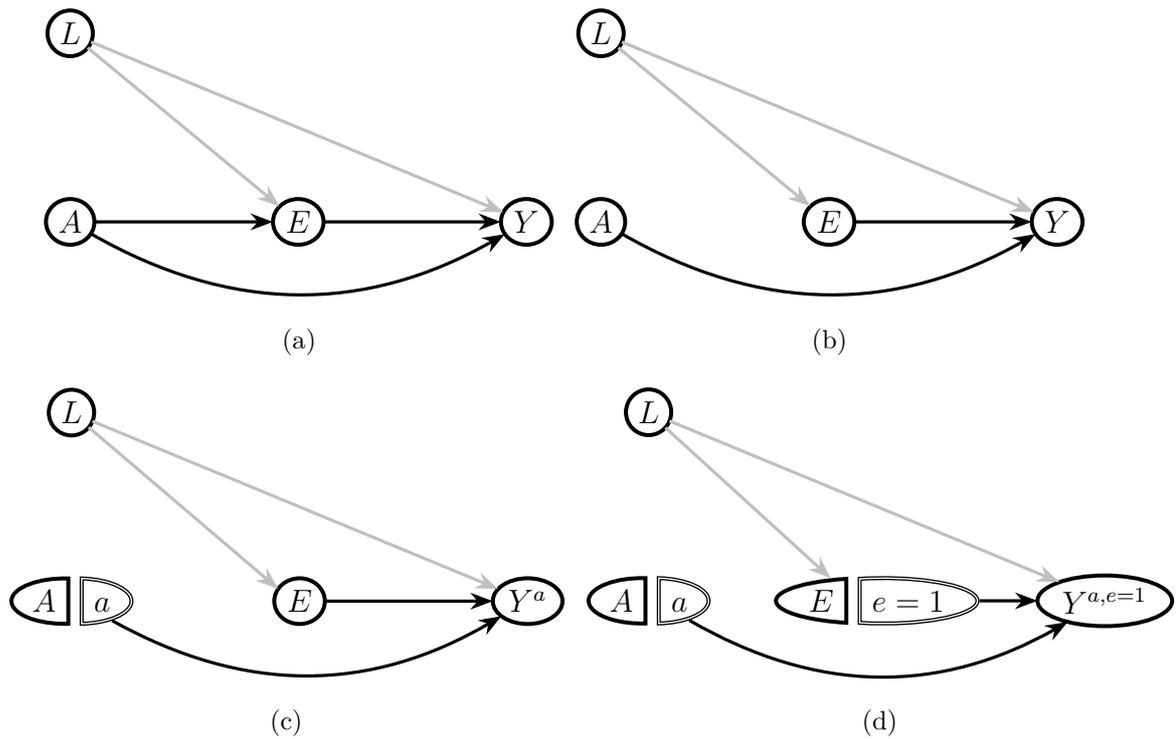
\begin{figure}
    \centering
    \subfloat[]{
\begin{tikzpicture}
\tikzset{line width=1.5pt, outer sep=0pt,
ell/.style={draw,fill=white, inner sep=2pt,
line width=1.5pt},
swig vsplit={gap=5pt,
inner line width right=0.5pt}};
\node[name=A,ell,  shape=ellipse] at (0,0) {$A$}  ;
\node[name=E,ell,  shape=ellipse] at (3,0) {$E$}  ;
    \node[name=Y,ell,  shape=ellipse] (Y) at (6,0) {$Y$};
    \node[name=L,ell,  shape=ellipse] at (0,2.5) {$L$};
\begin{scope}[>={Stealth[black]},
              every edge/.style={draw=black,very thick}]
	\path [->] (A) edge (E);
	\path [->] (A) edge[bend right] (Y);
    \path [->] (E) edge (Y);
    \path [->,>={Stealth[lightgray]}] (L) edge[lightgray]  (E);
    \path [->,>={Stealth[lightgray]}] (L) edge[lightgray]  (Y);
\end{scope}
\end{tikzpicture}
} 
    \subfloat[]{
\begin{tikzpicture}
\tikzset{line width=1.5pt, outer sep=0pt,
ell/.style={draw,fill=white, inner sep=2pt,
line width=1.5pt},
swig vsplit={gap=5pt,
inner line width right=0.5pt}};
\node[name=A,ell,  shape=ellipse] at (0,0) {$A$}  ;
\node[name=E,ell,  shape=ellipse] at (3,0) {$E$}  ;
    \node[name=Y,ell,  shape=ellipse] (Y) at (6,0) {$Y$};
    \node[name=L,ell,  shape=ellipse] at (0,2.5) {$L$};
\begin{scope}[>={Stealth[black]},
              every edge/.style={draw=black,very thick}]
	\path [->] (A) edge[bend right] (Y);
    \path [->] (E) edge (Y);
    \path [->,>={Stealth[lightgray]}] (L) edge[lightgray]  (E);
    \path [->,>={Stealth[lightgray]}] (L) edge[lightgray]  (Y);
\end{scope}
\end{tikzpicture}
} \\
\subfloat[]{
\begin{tikzpicture}
\tikzset{line width=1.5pt, outer sep=0pt,
ell/.style={draw,fill=white, inner sep=2pt,
line width=1.5pt},
swig vsplit={gap=5pt,
inner line width right=0.5pt}};
\node[name=E,ell,  shape=ellipse] at (3,0) {$E$}  ;
\node[name=A,shape=swig vsplit] at (0,0){
\nodepart{left}{$A$}
\nodepart{right}{$a$} };
    \node[name=Y,ell,  shape=ellipse] (Y) at (6,0) {$Y^{a}$};
    \node[name=L,ell,  shape=ellipse] at (0,2.5) {$L$};
\begin{scope}[>={Stealth[black]},
              every edge/.style={draw=black,very thick}]
    \path [->] (E) edge (Y);
    \path [->,>={Stealth[lightgray]}] (L) edge[lightgray]  (E.150);
    \path [->,>={Stealth[lightgray]}] (L) edge[lightgray]  (Y);
	\path [->] (A) edge[bend right] (Y);
\end{scope}
\end{tikzpicture}
}
\subfloat[]{
\begin{tikzpicture}
\tikzset{line width=1.5pt, outer sep=0pt,
ell/.style={draw,fill=white, inner sep=2pt,
line width=1.5pt},
swig vsplit={gap=5pt,
inner line width right=0.5pt}};
\node[name=E,shape=swig vsplit] at (3,0){
\nodepart{left}{$E$}
\nodepart{right}{$e=1$} };
\node[name=A,shape=swig vsplit] at (0,0){
\nodepart{left}{$A$}
\nodepart{right}{$a$} };
    \node[name=Y,ell,  shape=ellipse] (Y) at (6,0) {$Y^{a,e=1}$};
    \node[name=L,ell,  shape=ellipse] at (0,2.5) {$L$};
\begin{scope}[>={Stealth[black]},
              every edge/.style={draw=black,very thick}]
    \path [->] (E) edge (Y);
    \path [->,>={Stealth[lightgray]}] (L) edge[lightgray]  (E.150);
    \path [->,>={Stealth[lightgray]}] (L) edge[lightgray]  (Y);
	\path [->] (A) edge[bend right] (Y);
\end{scope}
\end{tikzpicture}
}
\\

    \caption{The Directed Acyclic Graph (DAG) in (a) describes a study where $A$ is randomly assigned. The DAG in (b) further encodes the no effect on exposure assumption, which is supposed to hold in a \textit{blinded} RCT. The graph in (c) is a Single World Intervention Graph (SWIG) where we have fixed the treatment to $a$. This SWIG can be used to study identifiability conditions for the Causal Effect Conditional on Exposure (CECE), which is identified even if $L$ is unmeasured. The SWIG in (d) describes interventions on both $A$ when $E$ ($e$ is fixed to 1), which allows us to study identifiability conditions for the CDE. Unlike the CECE, the Controlled Direct Effect (CDE) would require measurement of $L$.}
    \label{fig: swigs}
\end{figure}

\begin{figure}
\subfloat[]{
\begin{tikzpicture}
\tikzset{line width=1.5pt, outer sep=0pt,
ell/.style={draw,fill=white, inner sep=2pt,
line width=1.5pt},
swig vsplit={gap=5pt,
inner line width right=0.5pt}};
\node[name=A,shape=swig vsplit] at (-1,0){
\nodepart{left}{$A$}
\nodepart{right}{$a$} };
\node[name=E1,ell,  shape=ellipse] at (2,0) {$E_1^{a,c=0}$}  ;
\node[name=E2,ell,  shape=ellipse] at (8,0) {$E_2^{a,c=0}$}  ;
\node[name=Y1,ell,  shape=ellipse] at (5,0) {$Y_1^{a,c=0}$}  ;
\node[name=Y2,ell,  shape=ellipse] at (12,0) {$Y_2^{a,c=0}$}  ;
\node[name=L0,ell,  shape=ellipse] at (0,3) {$L_0$}  ;
\node[name=L1,ell,  shape=ellipse] at (5,3) {$L_1$}  ;
\node[name=C1,shape=swig vsplit] at (2,-3) { \nodepart{left}{$C_1^{a}$} 
\nodepart{right}{$c_1 = 0$}  };
\node[name=C2,shape=swig vsplit] at (8,-3){ \nodepart{left}{$C_2^{a,c=0}$}  
\nodepart{right}{$c_2 = 0$} };
\begin{scope}[>={Stealth[black]},
              every edge/.style={draw=black,very thick}]
	\path [->] (A) edge[bend right] (Y1);
	\path [->] (A) edge[bend right] (Y2);
	\path [->] (E1) edge (Y1);
	\path [->] (E1) edge[bend right] (E2);
	\path [->] (E1) edge[bend right] (Y2);
	\path [->] (E2) edge (Y2);
    \path [->,>={Stealth[lightgray]}] (L0) edge[lightgray]  (L1);
    \path [->,>={Stealth[lightgray]}] (L0) edge[lightgray]  (Y1);
    \path [->,>={Stealth[lightgray]}] (L0) edge[lightgray]  (Y2);
    \path [->,>={Stealth[lightgray]}] (L0) edge[lightgray]  (E1);
    \path [->,>={Stealth[lightgray]}] (L0) edge[lightgray]  (E2);
    \path [->,>={Stealth[lightgray]}] (L1) edge[lightgray]  (Y1);
    \path [->,>={Stealth[lightgray]}] (L1) edge[lightgray]  (Y2);
    \path [->,>={Stealth[lightgray]}] (L1) edge[lightgray]  (E2);
    \path [->,>={Stealth[lightgray]}] (A) edge[lightgray]  (C1);
    \path [->,>={Stealth[lightgray]}] (A) edge[lightgray]  (C2);
    \path [->,>={Stealth[lightgray]}] (E1) edge[lightgray]  (C2);
    \path [->,>={Stealth[lightgray]}] (Y1) edge[lightgray]  (C2);    
    \path [->,>={Stealth[lightgray]}] (C1) edge[lightgray]  (C2);    
\end{scope}
\end{tikzpicture}
} \\
    \caption{The Single World Intervention Graph (SWIG) shows a time-to-event setting where the Causal Effect Conditional on Exposure (CECE) is identified,  even if $L_0$ and $L_1$ are unmeasured. } 
    \label{fig:time to event}
\end{figure}
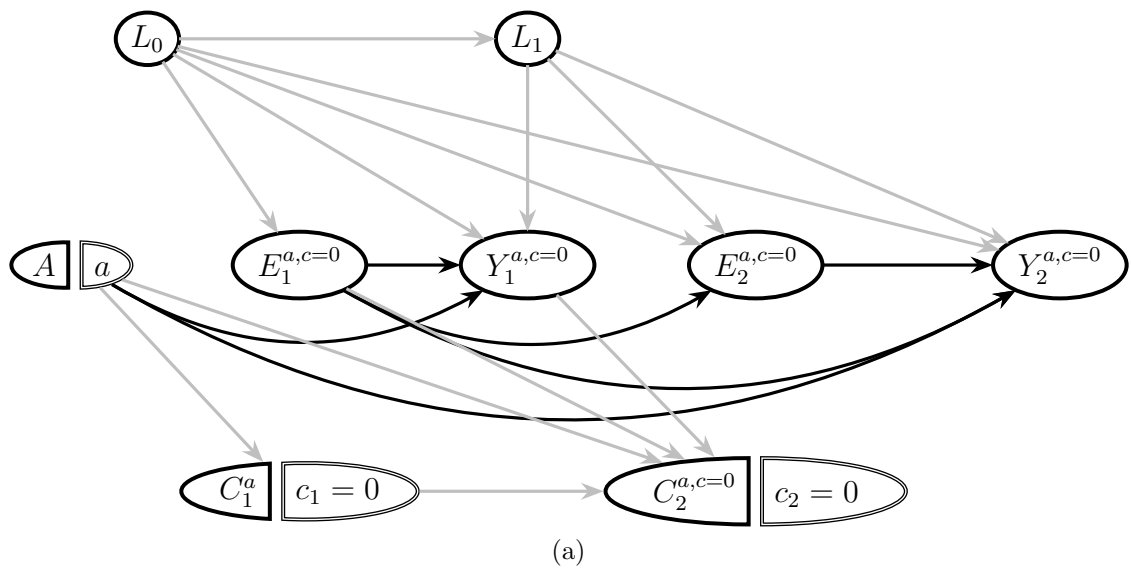

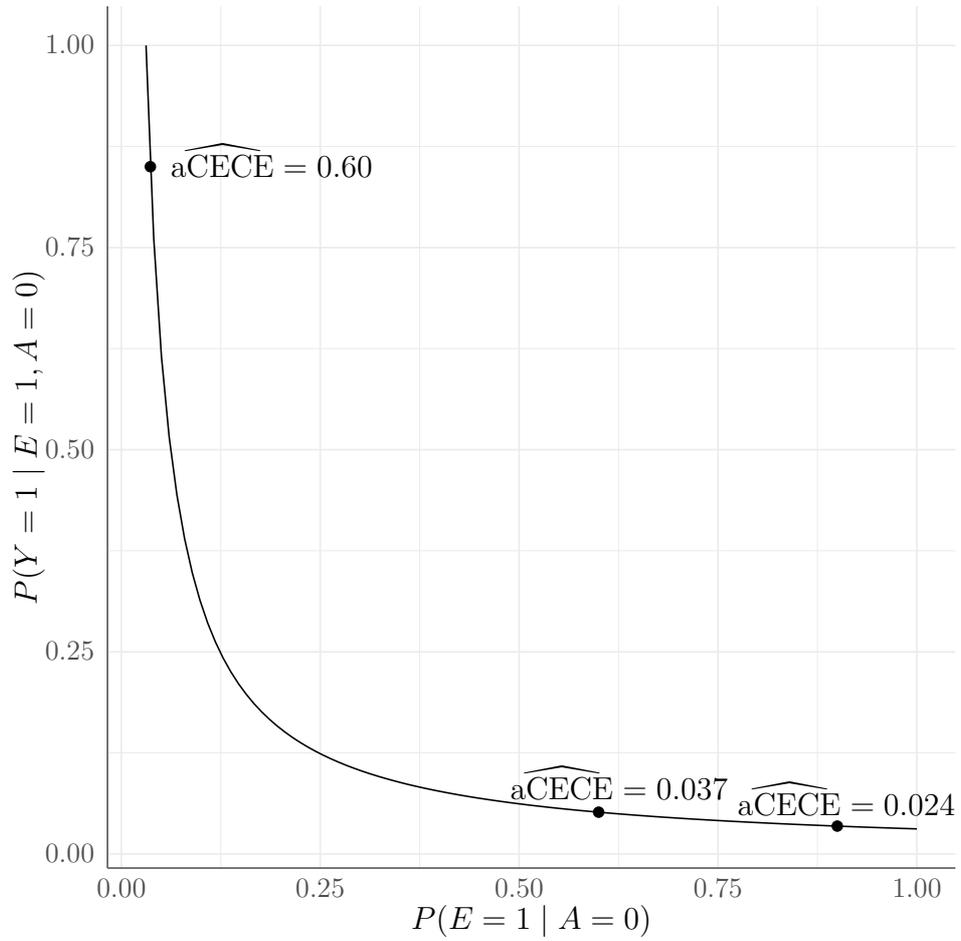
\begin{figure}
    \centering
\begin{tikzpicture}[x=1pt,y=1pt]
\definecolor{fillColor}{RGB}{255,255,255}
\path[use as bounding box,fill=fillColor,fill opacity=0.00] (0,0) rectangle (361.35,361.35);
\begin{scope}
\path[clip] ( 38.56, 30.69) rectangle (355.85,355.85);
\definecolor{drawColor}{gray}{0.92}

\path[draw=drawColor,line width= 0.3pt,line join=round] ( 38.56, 74.14) --
	(355.85, 74.14);

\path[draw=drawColor,line width= 0.3pt,line join=round] ( 38.56,150.41) --
	(355.85,150.41);

\path[draw=drawColor,line width= 0.3pt,line join=round] ( 38.56,226.67) --
	(355.85,226.67);

\path[draw=drawColor,line width= 0.3pt,line join=round] ( 38.56,302.94) --
	(355.85,302.94);

\path[draw=drawColor,line width= 0.3pt,line join=round] ( 80.96, 30.69) --
	( 80.96,355.85);

\path[draw=drawColor,line width= 0.3pt,line join=round] (155.38, 30.69) --
	(155.38,355.85);

\path[draw=drawColor,line width= 0.3pt,line join=round] (229.80, 30.69) --
	(229.80,355.85);

\path[draw=drawColor,line width= 0.3pt,line join=round] (304.22, 30.69) --
	(304.22,355.85);

\path[draw=drawColor,line width= 0.6pt,line join=round] ( 38.56, 36.01) --
	(355.85, 36.01);

\path[draw=drawColor,line width= 0.6pt,line join=round] ( 38.56,112.27) --
	(355.85,112.27);

\path[draw=drawColor,line width= 0.6pt,line join=round] ( 38.56,188.54) --
	(355.85,188.54);

\path[draw=drawColor,line width= 0.6pt,line join=round] ( 38.56,264.80) --
	(355.85,264.80);

\path[draw=drawColor,line width= 0.6pt,line join=round] ( 38.56,341.07) --
	(355.85,341.07);

\path[draw=drawColor,line width= 0.6pt,line join=round] ( 43.75, 30.69) --
	( 43.75,355.85);

\path[draw=drawColor,line width= 0.6pt,line join=round] (118.17, 30.69) --
	(118.17,355.85);

\path[draw=drawColor,line width= 0.6pt,line join=round] (192.59, 30.69) --
	(192.59,355.85);

\path[draw=drawColor,line width= 0.6pt,line join=round] (267.01, 30.69) --
	(267.01,355.85);

\path[draw=drawColor,line width= 0.6pt,line join=round] (341.43, 30.69) --
	(341.43,355.85);
\definecolor{drawColor}{RGB}{0,0,0}

\path[draw=drawColor,line width= 0.6pt,line join=round] ( 52.98,341.07) --
	( 55.86,268.42) --
	( 58.75,223.72) --
	( 61.63,193.44) --
	( 64.52,171.57) --
	( 67.40,155.04) --
	( 70.28,142.10) --
	( 73.17,131.70) --
	( 76.05,123.15) --
	( 78.94,116.01) --
	( 81.82,109.95) --
	( 84.71,104.74) --
	( 87.59,100.22) --
	( 90.48, 96.26) --
	( 93.36, 92.75) --
	( 96.25, 89.63) --
	( 99.13, 86.84) --
	(102.01, 84.33) --
	(104.90, 82.05) --
	(107.78, 79.97) --
	(110.67, 78.08) --
	(113.55, 76.34) --
	(116.44, 74.74) --
	(119.32, 73.26) --
	(122.21, 71.89) --
	(125.09, 70.62) --
	(127.97, 69.43) --
	(130.86, 68.33) --
	(133.74, 67.29) --
	(136.63, 66.32) --
	(139.51, 65.41) --
	(142.40, 64.55) --
	(145.28, 63.74) --
	(148.17, 62.97) --
	(151.05, 62.24) --
	(153.94, 61.56) --
	(156.82, 60.91) --
	(159.70, 60.29) --
	(162.59, 59.70) --
	(165.47, 59.14) --
	(168.36, 58.60) --
	(171.24, 58.09) --
	(174.13, 57.60) --
	(177.01, 57.13) --
	(179.90, 56.69) --
	(182.78, 56.26) --
	(185.66, 55.85) --
	(188.55, 55.45) --
	(191.43, 55.07) --
	(194.32, 54.71) --
	(197.20, 54.35) --
	(200.09, 54.02) --
	(202.97, 53.69) --
	(205.86, 53.37) --
	(208.74, 53.07) --
	(211.63, 52.78) --
	(214.51, 52.49) --
	(217.39, 52.22) --
	(220.28, 51.96) --
	(223.16, 51.70) --
	(226.05, 51.45) --
	(228.93, 51.21) --
	(231.82, 50.98) --
	(234.70, 50.75) --
	(237.59, 50.53) --
	(240.47, 50.32) --
	(243.35, 50.11) --
	(246.24, 49.91) --
	(249.12, 49.72) --
	(252.01, 49.53) --
	(254.89, 49.34) --
	(257.78, 49.16) --
	(260.66, 48.99) --
	(263.55, 48.82) --
	(266.43, 48.65) --
	(269.32, 48.49) --
	(272.20, 48.33) --
	(275.08, 48.18) --
	(277.97, 48.03) --
	(280.85, 47.88) --
	(283.74, 47.74) --
	(286.62, 47.60) --
	(289.51, 47.46) --
	(292.39, 47.33) --
	(295.28, 47.20) --
	(298.16, 47.07) --
	(301.04, 46.95) --
	(303.93, 46.83) --
	(306.81, 46.71) --
	(309.70, 46.59) --
	(312.58, 46.48) --
	(315.47, 46.37) --
	(318.35, 46.26) --
	(321.24, 46.15) --
	(324.12, 46.05) --
	(327.01, 45.95) --
	(329.89, 45.85) --
	(332.77, 45.75) --
	(335.66, 45.65) --
	(338.54, 45.56) --
	(341.43, 45.47);
\definecolor{fillColor}{RGB}{0,0,0}

\path[draw=drawColor,line width= 0.4pt,line join=round,line cap=round,fill=fillColor] (311.66, 46.52) circle (  1.96);

\path[draw=drawColor,line width= 0.4pt,line join=round,line cap=round,fill=fillColor] (222.36, 51.77) circle (  1.96);

\path[draw=drawColor,line width= 0.4pt,line join=round,line cap=round,fill=fillColor] ( 54.61,295.31) circle (  1.96);

\node[text=drawColor,anchor=base,inner sep=0pt, outer sep=0pt, scale=  1] at (315.04, 50.51) {$\widehat{\text{aCECE}} = 0.024$};

\node[text=drawColor,anchor=base,inner sep=0pt, outer sep=0pt, scale=  1] at (229.88, 56.65) {$\widehat{\text{aCECE}} = 0.037$};

\node[text=drawColor,anchor=base,inner sep=0pt, outer sep=0pt, scale=  1] at ( 99.97,291.51) {$\widehat{\text{aCECE}} = 0.60$};
\end{scope}
\begin{scope}
\path[clip] (  0.00,  0.00) rectangle (361.35,361.35);
\definecolor{drawColor}{RGB}{86,86,86}

\path[draw=drawColor,line width= 0.6pt,line join=round] ( 38.56, 30.69) --
	( 38.56,355.85);
\end{scope}
\begin{scope}
\path[clip] (  0.00,  0.00) rectangle (361.35,361.35);
\definecolor{drawColor}{gray}{0.30}

\node[text=drawColor,anchor=base east,inner sep=0pt, outer sep=0pt, scale=  0.88] at ( 33.61, 32.98) {0.00};

\node[text=drawColor,anchor=base east,inner sep=0pt, outer sep=0pt, scale=  0.88] at ( 33.61,109.24) {0.25};

\node[text=drawColor,anchor=base east,inner sep=0pt, outer sep=0pt, scale=  0.88] at ( 33.61,185.51) {0.50};

\node[text=drawColor,anchor=base east,inner sep=0pt, outer sep=0pt, scale=  0.88] at ( 33.61,261.77) {0.75};

\node[text=drawColor,anchor=base east,inner sep=0pt, outer sep=0pt, scale=  0.88] at ( 33.61,338.04) {1.00};
\end{scope}
\begin{scope}
\path[clip] (  0.00,  0.00) rectangle (361.35,361.35);
\definecolor{drawColor}{RGB}{86,86,86}

\path[draw=drawColor,line width= 0.6pt,line join=round] ( 38.56, 30.69) --
	(355.85, 30.69);
\end{scope}
\begin{scope}
\path[clip] (  0.00,  0.00) rectangle (361.35,361.35);
\definecolor{drawColor}{gray}{0.30}

\node[text=drawColor,anchor=base,inner sep=0pt, outer sep=0pt, scale=  0.88] at ( 43.75, 19.68) {0.00};

\node[text=drawColor,anchor=base,inner sep=0pt, outer sep=0pt, scale=  0.88] at (118.17, 19.68) {0.25};

\node[text=drawColor,anchor=base,inner sep=0pt, outer sep=0pt, scale=  0.88] at (192.59, 19.68) {0.50};

\node[text=drawColor,anchor=base,inner sep=0pt, outer sep=0pt, scale=  0.88] at (267.01, 19.68) {0.75};

\node[text=drawColor,anchor=base,inner sep=0pt, outer sep=0pt, scale=  0.88] at (341.43, 19.68) {1.00};
\end{scope}
\begin{scope}
\path[clip] (  0.00,  0.00) rectangle (361.35,361.35);
\definecolor{drawColor}{RGB}{0,0,0}

\node[text=drawColor,anchor=base,inner sep=0pt, outer sep=0pt, scale=  1] at (197.20,  7.64) {$P(E = 1 \mid A= 0)$};
\end{scope}
\begin{scope}
\path[clip] (  0.00,  0.00) rectangle (361.35,361.35);
\definecolor{drawColor}{RGB}{0,0,0}

\node[text=drawColor,rotate= 90.00,anchor=base,inner sep=0pt, outer sep=0pt, scale=  1] at ( 12.08,193.27) {$P(Y = 1 \mid E = 1, A = 0)$};
\end{scope}
\end{tikzpicture}
    \caption{Illustration of the relation between the sensitivity parameters $P(Y = 1 \mid  E = 1, A = 0) $ and $P(E = 1 \mid A = 0)$. Specifying one sensitivity parameter is sufficient to get a point estimate of the absolute Causal Effect Conditional on Exposure (CECE), as illustrated by three values of $\widehat{\text{aCECE}}$ in the figure.}
    \label{fig:sensitivity parameters}
\end{figure}
\clearpage

\renewcommand{\thesection}{Appendix \Alph{section}}
\renewcommand{\thesubsection}{\Alph{section}.\arabic{subsection}}
\renewcommand{\thesubsubsection}{\Alph{section}.\arabic{subsection}.\arabic{subsubsection}}

\section{Proofs}
\label{app sec: proofs}

\begin{proof}[Proof of Proposition \ref{thm: relative cdc}]

For any $a,a' \in \{0,1\}$, use laws of probability and  conditions \eqref{ass: no indirect vaccine effect} and \eqref{ass: cond trt ech}-\eqref{ass: exposure necessity} to express
\begin{align*}
   \mathbb{E} (Y \mid A = a) = &  \mathbb{E} (Y \mid E = 1, A = a) \underbrace{P(E = 1 \mid A = a)}_{= P(E = 1 \mid A = a')  }  + \underbrace { \mathbb{E} (Y \mid A = a, E = 0) P(E = 0 \mid A = a) }_{=0  } \nonumber \\
 =  &  \mathbb{E} (Y \mid E = 1, A = a)  P(E = 1 \mid A = a').
 \end{align*}
Thus,
\begin{align*}
    \frac{ \mathbb{E} (Y \mid A = 1)}{ \mathbb{E} (Y \mid A = 0)} & = \frac{ \mathbb{E} (Y \mid E = 1, A = 1) P(E = 1 \mid A = 1)}{ \mathbb{E} (Y \mid E = 1, A = 0) P(E = 1 \mid A = 0) } \\
    & = \frac{ \mathbb{E} (Y \mid E = 1, A = 1)}{ \mathbb{E} (Y \mid E = 1, A = 0)} \\
    & = \frac{\mathbb{E} (Y^{a=1} \mid E=1)}{ \mathbb{E} (Y^{a=0}  \mid E=1)},
\end{align*}
where the second and third line again follow due to assumption \eqref{ass: no indirect vaccine effect} and \eqref{ass: cond trt ech}-\eqref{ass: exposure necessity}.
\end{proof}

\begin{proof}[Proof of Proposition \ref{thm: absolute cdc}]
We first derive an upper bound. 
\begin{align}
 &{\mathbb{E} (Y^{a=0}  \mid E  = 1 ) } - {\mathbb{E} (Y^{a=1}  \mid E= 1 )} \nonumber \\
= &  \mathbb{E} (Y \mid E = 1, A = 0) - \mathbb{E} (Y \mid E = 1, A = 1)  \quad \text{due to} \ \eqref{ass: cond trt ech}-\eqref{ass: consistency}  \nonumber \\
= & \frac{\mathbb{E} (Y \mid  A = 0)   }{ P(E = 1 \mid A = 0)} -  \frac{\mathbb{E} (Y \mid A = 1)   }{ P(E = 1 \mid A = 1)  }  \quad  \text{due to}  \  \eqref{ass: exposure necessity} \ \text{and laws of prob.}   \nonumber \\
= &  \frac{\mathbb{E} (Y \mid  A = 0)   }{ P(E = 1 \mid A = 0)  } - \frac{\mathbb{E} (Y \mid A = 1)   }{ P(E = 1 \mid A = 0)  }  \quad \text{due to} \   \eqref{ass: no indirect vaccine effect}  \nonumber \\
\leq &  \frac{\mathbb{E} (Y \mid  A = 0)   }{ \mathbb{E} (Y \mid A = 0)  }  - \frac{\mathbb{E} (Y \mid A = 1)   }{ \mathbb{E} (Y \mid A = 0)  }   = 1 - \frac{\mathbb{E} (Y \mid A = 1)   }{  \mathbb{E} (Y \mid A = 0)  }. 
\label{eq: upper bound no indir}
\end{align}
The last line is an equality when $ (E=1 \iff Y = 1) \mid A = 0$.

A lower bound on the absolute CECE is given by
\begin{align}
 &{\mathbb{E} (Y^{a=0}  \mid E  = 1 ) } - {\mathbb{E} (Y^{a=1}  \mid E= 1 )} \nonumber \\
= &  \frac{\mathbb{E} (Y \mid  A = 0)   }{ P(E = 1 \mid A = 0)  } - \frac{\mathbb{E} (Y \mid A = 1)   }{ P(E = 1 \mid A = 0)  }   \nonumber \\
\geq &  {\mathbb{E} (Y \mid  A = 0)   }  - {\mathbb{E} (Y \mid A = 1)   } . 
\label{eq: lower bound no indir}
\end{align}
The last line in \eqref{eq: lower bound no indir} is an equality when $P(E = 1 ) = 1$. 
\end{proof}

\begin{proof}[Proof of Proposition \ref{thm: conditional per exposure eff}]
\begin{align}
& \frac{\mathbb{E} (Y^{a=1,e=1} \mid L  ) }{ \mathbb{E} (Y^{a=0,e=1} \mid L)  } = \frac{ \mathbb{E} (Y \mid E = 1, A = 1, L ) }{ \mathbb{E} (Y \mid E = 1, A = 0, L )  } \quad \text{due to }  \eqref{ass: cond exp ech}-\eqref{ass: exposure con}  \nonumber \\
= & \frac{ \mathbb{E} (Y \mid  A = 1, L )}{ \mathbb{E} (Y \mid  A = 0, L ) }, \nonumber
\end{align}
where the last equality follows from 
$$
     \mathbb{E} (Y \mid E = 1, A = a, L  )=\frac{ \mathbb{E} (Y \mid  A = a, L )}{ P(E = 1 \mid A = a, L ) }  = \frac{ \mathbb{E} (Y \mid  A = a, L )}{ P(E = 1 \mid A = a', L ) },
    $$
using \eqref{ass: exposure necessity} and \eqref{ass: no indirect vaccine effect}, similarly to the proof of Proposition \eqref{thm: relative cdc}.
\end{proof}

\begin{proof}[Proof of Proposition \ref{corollary: cece per exp}]
The result follows from including $L$ in the conditioning set in all the derivations of Proposition 1, which then gives the same identification result as in Proposition \ref{thm: conditional per exposure eff}.
\end{proof}


\section{Time-to-events and censoring}
\label{sec app: time-to-event}
We re-introduce the terminology from Section \nameref{sec: time to event}. Let $Y_k$ and $E_k$ be time-to-event variables indicating whether an individual has experienced the event by time $k$ ($Y_k = 1$) and being exposed by time $k$ ($E_k = 1$), respectively. Let $C_{k}$ denote loss to follow-up (censoring) by interval $k > 0$, and we define the temporal (and topological) order $(C_{k},E_{k},Y_{k})$ in each interval $k > 0$. Suppose we are interested in outcomes in time intervals $k=0,\dots, K$. We adopt the convention that random variables with a negative subscript are equal to 0 (e.g., $Y_{-1} \equiv 0$). 

Let the history of a random variable be denoted by a check symbol, e.g. $\check{Y}_{k}=(Y_{0},Y_{1},...,Y_{k})$ is the history of the event of interest through interval $k$. Further, let the future of a random variable through $K$ be denoted by an underline, e.g. $\underline{Y}_{k}=(Y_{k},Y_{k+1},...,Y_{K})$.

Consider now classical identifiability conditions for causal effects in time-to-event settings, which are just extensions of \eqref{ass: cond trt ech}-\eqref{ass: consistency}. 

\begin{assumption}[Treatment exchangeability]
\begin{align}
& \check{Y}_{K}^{a, c=0}, \check{E}_{K}^{a, c=0} \independent A, \label{ass: time ech} \\
&\underline{Y}^{a, c=0}_{k} \independent C_{k} \mid Y_{k-1} =  C_{k-1} = 0, A = a. \label{ass: independence censoring}
\end{align} 
 \end{assumption}
 Condition \eqref{ass: time ech} holds when $A$ is randomly assigned. Condition \eqref{ass: independence censoring} requires that losses to follow-up are independent of future counterfactual events, given the measured past; this assumption, which corresponds to classical independent censoring assumptions, does not hold by design in a randomized trial, as losses to follow-up are not randomly assigned in practice. The treatment exchangeability conditions are satisfied in the SWIG in Figure \ref{fig:time to event}.

\begin{assumption}[Positivity]
\begin{align}
& P(A =a ) > 0 \quad  \forall a \in \{0,1\}  \label{ass: time pos} \\
& P( Y_k = 0,  C_k = 0,A=a) > 0  \implies \nonumber \\ 
& \quad  P(C_{k+1} = 0 \mid Y_k = 0, C_k = 0, A=a)>0 \  \label{eq: time pos 2} ,
\end{align} 
for all $ a \in \{0,1\}$ and $k < K$.
 \end{assumption}  The positivity conditions require that for any possible history of treatment assignment and covariates among those who are event-free and uncensored at $k$, some subjects will remain uncensored at the next time $k+1$.
 
 \begin{assumption}[Consistency]
\begin{align}
  & \text{if } A=a \text{ and } C_{k} = 0, \nonumber \\
  & \text{then } \check{Y}_{k} = \check{Y}^{a,  c=0}_{k}, \check{E}_{k} = \check{E}^{a,  c=0}_{k} \,
  \label{ass: time consistency}
\end{align} 
for all $ a \in \{0,1\}$ and $k \leq K$.
 \end{assumption}
Consistency holds if any individual who has data history consistent with the intervention under a counterfactual scenario, would have observed outcomes that are equal to the counterfactual outcomes. 


Besides the classical identifiability conditions, we introduce the following conditions, which generalize exposure necessity \eqref{ass: exposure necessity} and the no effect on exposure assumption \eqref{ass: no indirect vaccine effect} from the main text.

\begin{assumption}[Time-varying exposure necessity]
\begin{align}
E_k^{a,c=0} = 0 \implies Y_k^{a,c=0} = 0 ,
\label{cond: time-var exposure necessity} 
\end{align}
for all $ a \in \{0,1\}$ and $k \leq K$.
\end{assumption}

Like \eqref{ass: exposure necessity}, the assumption of time-varying exposure necessity states that the outcome can only happen in individuals who have been exposed to the virus. By definition, $E^{a,c=0}_{k-1} = 1 \implies E^{a,c=0}_{k} = 1$, so the time-varying nature of exposure and the outcome should not make this assumption less justifiable.


\begin{assumption}[No effect on exposure]
\begin{align}
    E^{a=0,c=0}_{k}   =
    E^{a=1,c=0}_{k} , \ \label{ass no exposure eff time-var} 
\end{align}
for all $k \leq K$.
\end{assumption}
 
This assumption says that the risk of exposure by any time $k$ is the same among treated and untreated. Consider a situation in which a vaccine $A$ prevents or delays the outcome $Y$. Under blinding, condition \eqref{ass no exposure eff time-var} would still hold because prior infection would be the only thing preventing future exposure, but under \eqref{cond: time-var exposure necessity}, anyone with the outcome would have already been exposed. However, we must assume that blinding continues to be successful; that is, this assumption would be violated if over time individuals notice that they are not getting infected after the same level of exposure as people around them, and therefore conclude that they have been vaccinated and change behavior.


Under these conditions we sketch a proof for Proposition \ref{thm: rel CECE time var}.

\begin{proof}[Sketch of proof of Proposition \ref{thm: rel CECE time var}.]
We can invoke \eqref{cond: time-var exposure necessity}-\eqref{ass no exposure eff time-var} to find that
\begin{align*}
    & \frac{ \mathbb{E} (Y_{k}^{a=1,c=0})}{ \mathbb{E} (Y_{k}^{a=0,c=0})} = \frac{ \mathbb{E} (E_{k}^{a=1,c=0} Y_{k}^{a=1,c=0}  )}{ \mathbb{E} (E_{k}^{a=0,c=0} Y_{k}^{a=0,c=0})} \\
    =&  \frac{ \mathbb{E} ( Y_{k}^{a=1,c=0} \mid E_{k}^{a=1,c=0} =1 ) \mathbb{E}(E_{k}^{a=1,c=0}) }{ \mathbb{E} ( Y_{k}^{a=0,c=0} \mid E_{k}^{a=0,c=0} = 1) \mathbb{E}(E_{k}^{a=0,c=0})} = \frac{ \mathbb{E} ( Y_{k}^{a=1,c=0} \mid E_{k}^{a=1,c=0} =1 )  }{ \mathbb{E} ( Y_{k}^{a=0,c=0} \mid E_{k}^{a=0,c=0} = 1)  },
\end{align*}
where we used exposure necessity in the first equality, laws of probability in the second equality and the last equality follows because
 $\mathbb{E}(E_{k}^{a=0,c=0}) = \mathbb{E}(E_{k}^{a=1,c=0})$ under \eqref{ass no exposure eff time-var}.

 Then, using treatment exchangeability, consistency and positivity, it follows that $\mathbb{E} (Y_{k}^{a,c=0})$ can be expressed in terms of the cumulative incidence function at $k$, $\mu_k (a)$.
 
 The proof for the additive CECE follows the same structure as the proof of Proposition \ref{thm: absolute cdc}.
\end{proof}

\section{Parallel to risk ratio under perfect specificity}\label{sec app:specificity}

\begin{figure}[h]
\subfloat[]{
\begin{tikzpicture}
\tikzset{line width=1.5pt, outer sep=0pt,
ell/.style={draw,fill=white, inner sep=2pt,
line width=1.5pt},
swig vsplit={gap=5pt,
inner line width right=0.5pt}};
\node[name=A,ell,  shape=ellipse] at (0,0) {$A$}  ;
\node[name=Y,ell,  shape=ellipse] (Y) at (3,0) {$Y$};
\node[name=Ys,ell,  shape=ellipse] at (6,0) {$Y^*$}  ;
\begin{scope}[>={Stealth},
              every edge/.style={draw}]
	\path [->, very thick] (A) edge (Y);
    \path [->, ultra thick] (Y) edge (Ys);
\end{scope}
\end{tikzpicture}
} 
\subfloat[]{
\begin{tikzpicture}
\tikzset{line width=1.5pt, outer sep=0pt,
ell/.style={draw,fill=white, inner sep=2pt,
line width=1.5pt},
swig vsplit={gap=5pt,
inner line width right=0.5pt}};
\node[name=A,ell,  shape=ellipse] at (0,0) {$A$}  ;
\node[name=E,ell,  shape=ellipse] at (3,0) {$E$}  ;
    \node[name=Y,ell,  shape=ellipse] (Y) at (6,0) {$Y$};
\begin{scope}[>={Stealth},
              every edge/.style={draw}]
	\path [->, very thick] (A) edge[bend right] (Y);
    \path [->, ultra thick] (E) edge (Y);
\end{scope}
\end{tikzpicture}
}
\caption{Simplified DAGs demonstrating the parallels described in e\ref{sec app:specificity}. (a) Non-differential misclassification of the outcome. The assumption that outcome misclassification doesn't depend on exposure results in $A \independent Y^* \mid Y$.  The heavier arrow from $Y$ to $Y^*$ represents the perfect specificity assumption: $Y = 0 \implies Y^* = 0$. (b) The setting from the main text (simplified to remove common causes of $E$ and $Y$). The no effect on exposure assumption results in $A \independent E$. The heavier arrow from $E$ to $Y$ represents the exposure necessity assumption: $E = 0 \implies Y = 0$.}
\label{fig: specificity dags}
\end{figure}
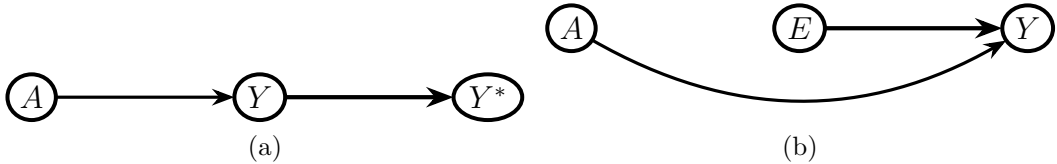

A well-known result in epidemiology is the fact that under so-called non-differential misclassification of the outcome with perfect specificity, the exposure-outcome risk ratio is unbiased, although the risk difference is not. For example, in the setting of possibly incomplete disease ascertainment in exposed and unexposed cohorts, Lawrence and Greenwald described how a screening program could be implemented to remove false positive cases, resulting in an unbiased risk ratio \cite{lawrence1977epidemiologic}. The requirement of perfect specificity parallels our exposure necessity assumption, and that of non-differential misclassification parallels our assumption of no effect on exposure. We demonstrate these parallels with the DAGs in Figure \ref{fig: specificity dags}. Each has one partially deterministic arrow and one independence assumption, though the causal structures differ. The partially deterministic arrow and the independence assumption allow in each case for an unbiased ratio measure, as we demonstrate in the following derivation. Take $Y$ to be a binary outcome and $A$ any exposure of interest (also binary for simplicity). We denote a misclassified version of the outcome with $Y^*$. Then we have for the misclassification setting that
\begin{align*}
    & \frac{P(Y^* = 1 \mid A = 1)}{P(Y^* = 1 \mid A = 0)} \\
    & = \frac{P(Y^* = 1 \mid A = 1, Y = 1) P(Y = 1 \mid A = 1) + P(Y^* = 1 \mid A = 1, Y = 0) P(Y = 0 \mid A = 1)}{P(Y^* = 1 \mid A = 0, Y = 1) P(Y = 1 \mid A = 0) + P(Y^* = 1 \mid A = 0, Y = 0) P(Y = 0 \mid A = 0)} \\
    & = \frac{P(Y^* = 1 \mid A = 1, Y = 1) P(Y = 1 \mid A = 1) + 0\times P(Y = 0 \mid A = 1)}{P(Y^* = 1 \mid A = 0, Y = 1) P(Y = 1 \mid A = 0) + 0\times P(Y = 0 \mid A = 0)}\\
    & = \frac{P(Y^* = 1 \mid  Y = 1) P(Y = 1 \mid A = 1) }{P(Y^* = 1 \mid  Y = 1) P(Y = 1 \mid A = 0)} = \frac{P(Y = 1 \mid A = 1)}{P(Y = 1 \mid A = 0)},
\end{align*}
and a parallel derivation of our Proposition \ref{thm: relative cdc}, that is,
\begin{align*}
    & \frac{P(Y = 1 \mid A = 1)}{P(Y = 1 \mid A = 0)} \\
    & = \frac{P(Y = 1 \mid A = 1, E = 1) P(E = 1 \mid A = 1) + P(Y = 1 \mid A = 1, E = 0) P(E = 0 \mid A = 1)}{P(Y = 1 \mid A = 0, E = 1) P(E = 1 \mid A = 0) + P(Y = 1 \mid A = 0, E = 0) P(E = 0 \mid A = 0)}\\
    & = \frac{P(Y = 1 \mid A = 1, E = 1) P(E = 1 \mid A = 1) + 0\times P(E = 0 \mid A = 1)}{P(Y = 1 \mid A = 0, E = 1) P(E = 1 \mid A = 0) + 0\times P(E = 0 \mid A = 0)} \\
     & = \frac{P(Y = 1 \mid A = 1, E = 1) P(E = 1)}{P(Y = 1 \mid A = 0, E = 1) P(E = 1 )} = \frac{P(Y = 1 \mid A = 1, E = 1)}{P(Y = 1 \mid A = 0, E = 1)},
\end{align*}

where the second equality uses the appropriate partially deterministic arrow assumption and the third equality the appropriate independence assumption.

\section{Identification of the CDE \eqref{eq: cdce}}
\label{sec: id cde}
Identification results for the CDE are well-established when the exposure $E$ is measured. To motivate our new result for settings where $E$ is unmeasured, we first present these conventional identifiability conditions. 
\begin{assumption}[Exposure exchangeability]
\begin{align}
& E^{a} \independent A \mid L, \ Y^{a,e=1} \independent A \mid L \ \text{ and } \ Y^{a,e=1} \independent E^{a} \mid L,A=a. \label{ass: cond exp ech}
\end{align} 
 \end{assumption}
Conditions \eqref{ass: cond exp ech} are classical exchangeability condition, analogous to assumptions that are typically implemented to identify per protocol effects in trials, causal effects from observational data, and mediation effects. This assumption is stronger than exchangeability assumption \eqref{ass: cond trt ech}. In particular, \eqref{ass: cond exp ech} does not hold unless we measure common causes of $Y$ and $E$, as illustrated in the SWIG in Figure \ref{fig: swigs}d.

 \begin{assumption}[Exposure positivity]
\begin{align}
& P(A =a, E = 1 \mid L) > 0 \quad   \ \text{ for all } a\in \{0,1\} \text{ w.p.1. } \label{ass: exposure pos}
\end{align} 
\end{assumption}

 \begin{assumption}[Exposure consistency]
\begin{align}
& \text{ If }    A=a \text{ then } {E} = {E}^{a}. \text{ If }    A=a \text{ and } E=1  \text{ then } {Y} = {Y}^{a,e=1},  \label{ass: exposure con}
\end{align} 
 for all  $a\in \{0,1\}$ .
 \end{assumption}

When we impose conditions \eqref{ass: cond exp ech}-\eqref{ass: exposure con}, the CDE can be expressed as a function of factual variables,
$$
\mathbb{E} (Y^{a,e=1}) = \mathbb{E} \{ \mathbb{E} (Y \mid E = 1, A = a, L) \}.
$$
However, because we do not measure $E$, it is not possible to identify the CDE from our observed data; we cannot identify the term $\mathbb{E} (Y \mid E = 1, A = a, L) $ without measuring $E$. In particular, the absolute CDE cannot be identified unless we measure $E$. Nevertheless, our next proposition shows that the relative CDE conditional on $L$ is point identified. 

\begin{proposition}[CDE conditional on $L$]
\label{thm: conditional per exposure eff}
Under conditions \eqref{ass: no indirect vaccine effect}, \eqref{ass: exposure necessity} and \eqref{ass: cond exp ech}-\eqref{ass: exposure con}, the relative CDE conditional on the baseline covariate $L$ is 
$$
 \frac{\mathbb{E} (Y^{a=1,e=1} \mid L  ) }{ \mathbb{E} (Y^{a=0,e=1} \mid L)  }  = \frac{ \mathbb{E} (Y \mid  A = 1, L )}{ \mathbb{E} (Y \mid  A = 0, L ) }. $$
\end{proposition}
The proof is given in \ref{app sec: proofs}. The following proposition relates the CECE within a subpopulation defined by $L$ and the CDE. 

\begin{proposition}[CECE and CDE conditional on $L$]
\label{corollary: cece per exp}
 Under conditions \eqref{ass: no indirect vaccine effect}, \eqref{ass: cond trt ech}-\eqref{ass: exposure necessity} and \eqref{ass: cond exp ech}-\eqref{ass: exposure con}, the relative CECE given $L=l$ and the relative CDE conditional on the baseline covariate $L=l$ are equal, that is, 
 $$
 \frac{\mathbb{E} (Y^{a=1}  \mid E^{a=1}  = 1,  L=l ) }{\mathbb{E} (Y^{a=0}  \mid E^{a=0}= 1 , L=l )} = \frac{\mathbb{E} (Y^{a=1,e=1} \mid L=l  ) }{ \mathbb{E} (Y^{a=0,e=1} \mid L=l)  }  = \frac{ \mathbb{E} (Y \mid  A = 1, L=l )}{ \mathbb{E} (Y \mid  A = 0, L=l ) }. 
    $$
\end{proposition}

It is crucial that the covariate vector $L$ in Proposition \ref{thm: conditional per exposure eff} and Proposition \ref{corollary: cece per exp} is sufficient to adjust for confounding, i.e.\ to ensure that exposure exchangeabillity \eqref{ass: cond exp ech} holds. Thus, identification of the conditional CDEs requires stronger assumptions compared to identification of the CECE. Although the conditional CECE can be defined and estimated within any set of baseline covariates, it is only interpretable as a conditional CDE when that set of covariates consists of those sufficient to adjust for confounding.

\subsection*{The marginal CDE is not point identified}
Whereas the conditional relative CDE can be point identified under \eqref{ass: no indirect vaccine effect}, \eqref{ass: exposure necessity} and \eqref{ass: cond exp ech}-\eqref{ass: exposure con}, the marginal relative CDE is not identified without additional assumptions. To see this, consider a binary outcome $Y \in \{0,1\}$. The marginal relative CDE can be expressed as a weighted average of conditional relative CDEs,\footnote{Following the collapsibility results for relative risks in e.g.\ Huitfeldt et al \cite{huitfeldt2019collapsibility}.}
\begin{align}
 \frac{\mathbb{E} (Y^{a=1,e=1}   ) }{ \mathbb{E} (Y^{a=0,e=1} )  }  = &  \sum_l \frac{\mathbb{E} (Y^{a=1,e=1} \mid L = l  ) }{ \mathbb{E} (Y^{a=0,e=1} \mid L = l )  } P ( L = l  \mid Y^{a=0,e=1} = 1 )  \nonumber \\ 
= &  \sum_l \frac{ \mathbb{E} (Y \mid  A = 1, L = l)}{ \mathbb{E} (Y \mid  A = 0, L = l ) }   P ( L = l  \mid Y^{a=0,e=1} = 1 ).
\end{align}
Using laws of probability and \eqref{ass: cond exp ech}-\eqref{ass: exposure con}, $P ( L = l  \mid Y^{a=0,e=1} = 1 )$ can be written as 
\begin{align*}
    P ( L = l  \mid Y^{a=0,e=1} = 1 ) & = \frac{ P (  Y^{a=0,e=1} = 1 \mid L = l ) P ( L = l)  }{P(Y^{a=0,e=1} = 1)} \\
    & = \frac{ P (  Y = 1 \mid E =1, A = 0, L = l ) P ( L = l)  }{ \sum_l P(Y= 1 \mid E = 1, A = 0 ,L = l) P(L=l)}, 
\end{align*}
which depends on probabilities conditional on $E=1$ that are not estimable from observed data. However, we can point-identify the marginal CDE under the additional strong assumption that $\mathbb{E} (Y^{a=0,e=1}) = 1$, that is, the exposure deterministically causes the outcome if untreated. Then,
$    P ( L = l  \mid Y^{a=0,e=1} = 1 )  = P ( L = l) $,
and thus the marginal relative CDE is point identified by
$$
 \frac{\mathbb{E} (Y^{a=1,e=1}   ) }{ \mathbb{E} (Y^{a=0,e=1} )  } 
=  \sum_l \frac{ \mathbb{E} (Y \mid  A = 1, L = l)}{ \mathbb{E} (Y \mid  A = 0, L = l ) }   P ( L = l  ).
$$
The marginal absolute PPE is point identified as 
$$
\mathbb{E} (Y^{a=1,e=0}) - \mathbb{E} (Y^{a=1,e=1}) = 1 -  \frac{\mathbb{E} (Y^{a=1,e=1}   ) }{ \mathbb{E} (Y^{a=0,e=1} )  }  = 1 - \sum_l \frac{ \mathbb{E} (Y \mid  A = 1, L = l)}{ \mathbb{E} (Y \mid  A = 0, L = l ) }   P ( L = l  ).
$$

\section{Sensitivity analyses}
\label{sec app: sensitivity analysis}


\begin{proof}[Proof of Proposition \ref{thm: sensitivity motivation} from Section "\nameref{sec: id external data}". ]

We follow the same strategy as for the lower bound \eqref{eq: upper bound no indir} in the main text.

\begin{align*}
 &{\mathbb{E} (Y^{a=1}  \mid E^{a=1}  = 1 ) } - {\mathbb{E} (Y^{a=0}  \mid E^{a=0}= 1 )} \nonumber \\
= &  \mathbb{E} (Y \mid E = 1, A = 0) - \mathbb{E} (Y \mid E = 1, A = 1)  \quad \text{due to} \ \eqref{ass: cond trt ech}-\eqref{ass: consistency}  \nonumber \\
= & \ 
\frac{\mathbb{E} (EY \mid  A = 0)   }{ P(E = 1 \mid A = 0)  }  - \frac{\mathbb{E} (EY \mid A = 1)   }{ P(E = 1 \mid A = 1)  }  \quad  \text{(laws of prob.)}  \nonumber \\
= & \frac{\mathbb{E} (Y \mid  A = 0)   }{ P(E = 1 \mid A = 0)} -  \frac{\mathbb{E} (Y \mid A = 1)   }{ P(E = 1 \mid A = 1)  }  \quad  \text{due to}  \  \eqref{ass: exposure necessity}   \nonumber \\
= &  \frac{\mathbb{E} (Y \mid  A = 0)   }{ P(E = 1 \mid A = 0)  } - \frac{\mathbb{E} (Y \mid A = 1)   }{ P(E = 1 \mid A = 0)  }  \quad \text{due to} \   \eqref{ass: no indirect vaccine effect}  \nonumber \\
= &  \frac{\mathbb{E} (Y \mid  A = 0)  \mathbb{E} (Y  \mid E = 1, A = 0) }{ \mathbb{E} (Y \mid A = 0)  } - \frac{\mathbb{E} (Y \mid A = 1)  \mathbb{E} (Y  \mid E = 1, A = 0) }{ \mathbb{E} (Y \mid A = 0)  }   \nonumber \\  
= &  \mathbb{E} (Y  \mid E = 1, A = 0)  \left( 1 - \frac{\mathbb{E} (Y \mid A = 1)  }{ \mathbb{E} (Y \mid A = 0)  } \right) ,
\end{align*}
where we used exposure necessity \eqref{ass: exposure necessity} in the 5th equality, which implies that
$
P(Y = 1 \mid E = 1, A = 0) P( E = 1 \mid A = 0) = 
P(Y = 1 \mid  A = 0)
$ and that $\mathbb{E} (Y \mid A = 0)  \geq \mathbb{E} (Y \mid A = 1) $.
\end{proof}
Proposition \ref{thm: sensitivity motivation} motivates a sensitivity analysis and/or use of data from external sources; the investigator can include their background knowledge on $P(Y = 1 \mid E = 1, A = 0) $ -- the probability of experiencing the outcome given exposure in the unvaccinated -- along with the observed data on $ \mathbb{E} (Y  \mid  A = a) $ to point identify the absolute CECE. 

The 4th line of the proof of Proposition \ref{thm: sensitivity motivation} motivates an alternative sensitivity analysis: the investigator can specify the marginal risk of being exposed to the infectious agent given no treatment, that is, $ P(E = 1 \mid A = 0)  $, and then point identify the risk difference.

\section{R code to compute bounds for the aCECE and create Figure 3}

\setstretch{1}
\begin{verbatim}
\footnotesize
# install.packages(c("ggplot2", "ggrepel"))
library(ggplot2)

## uncomment these lines and dev.off() at the bottom to recreate the
## figure using tikz
## install.packages("tikzDevice")
# library(grid)
# library(tikzDevice)
# 
# tikz(file = "sensitivity_params.tex", 
#      standAlone = FALSE,
#      width = 5, 
#      height = 5
# )

## Observed values 
# P(Y = 1 | A = 1)
pYA1 <- 0.009
# P(Y = 1 | A = 0)
pYA0 <- 0.031

# function to compute the bounds of the aCECE
calc_aCECE <- function(pYA1, pYA0, pE = 1, pYE1A0 = 1) {
  lower_bound <- (pYA0 - pYA1) / pE
  upper_bound <- (1 - pYA1 / pYA0) * pYE1A0
  c(lower_bound = lower_bound, upper_bound = upper_bound)
}

# calculate the bounds using the observed values set above
calc_aCECE(pYA1 = pYA1, pYA0 = pYA0)

# given a certain Pr(Y = 1 | E = 1, A = 0), we can compute Pr(E = 1 | A = 0)
# from the observed P(Y = 1 | A = 0) and P(Y = 1 | A = 1) and vice versa
convert_params <- function(pYE1A0 = NULL, pE = NULL, pYA1, pYA0) {
  if (!is.null(pYE1A0) & !is.null(pE)) stop("Please specify only one of pYE1A0 
  and pE")
  if (is.null(pYE1A0) & is.null(pE)) stop("Either pYE1A0 or pE must be specified")
  if (!is.null(pYE1A0)) {
    pE <- -((pYA1 - pYA0) / (1 - (pYA1 / pYA0))) / pYE1A0
    return(c(pE = pE))
  } else {
    pYE1A0 <- -((pYA1 - pYA0) / (1 - (pYA1 / pYA0))) / pE
    return(c(pYE1A0 = pYE1A0))
  }
}

# confirm that we can go back and forth
convert_params(pYE1A0 = 0.05, pYA1 = pYA1, pYA0 = pYA0)
convert_params(pE = 0.62, pYA1 = pYA1, pYA0 = pYA0)

# as soon as we specify one or the other of Pr(E = 1 | A = 0) or 
# Pr(Y = 1 | E = 1, A = 0), the bounds are the same 
# and the aCECE is point-identified
# here we are specifying Pr(Y = 1 | E = 1, A = 0) = 0.85 and using that to 
# compute Pr(E = 1 | A = 0)
calc_aCECE(pYA1 = pYA1, pYA0 = pYA0, 
           pE = convert_params(pYE1A0 = 0.85, pYA1 = pYA1, pYA0 = pYA0), 
           pYE1A0 = 0.85)

# specify values to compute aCECE for
pE <- c(0.9, 
        0.6, 
        convert_params(pYE1A0 = 0.85, pYA1 = pYA1, pYA0 = pYA0))
pYE1A0 <- c(convert_params(pE = 0.9, pYA1 = pYA1, pYA0 = pYA0), 
            convert_params(pE = 0.6, pYA1 = pYA1, pYA0 = pYA0), 
            0.85)

# apply the function to each pair of values
aCECE <- mapply(FUN = calc_aCECE, 
                pE = pE, pYE1A0 = pYE1A0, 
                # the other parameters stay the same
                MoreArgs = list(pYA1 = pYA1, pYA0 = pYA0))

# the lower and upper bounds are the same
aCECE

# create dataframe of points to graph
points_to_highlight <- data.frame(x = pE, y = pYE1A0, aCECE = aCECE[1,],
                                  lbl = paste0("aCECE = ",
                                  scales::number(aCECE[1,])))

# plot the range of possible pairs of 
# Pr(E = 1 | A = 0) and Pr(Y = 1 | E = 1, A = 0)
# with those values annotated with the corresponding aCECE
fig <- ggplot(data = points_to_highlight, aes(x, y, label = lbl)) +
  xlim(pYA0, 1) +
  geom_function(fun = convert_params, args = list(pYA1 = pYA1, pYA0 = pYA0)) +
  labs(x = "$\\Pr(E = 1 \\mid A= 0)$", y = "$\\Pr(Y = 1 \\mid E = 1, A = 0)$") +
  geom_point() + 
  ggrepel::geom_text_repel() + 
  theme_minimal() +
  theme(axis.line.y.left = element_line(color = "#565656"),
        axis.line.x.bottom = element_line(color = "#565656"))

print(fig)     

# dev.off()

\end{verbatim}

\end{document}